\documentclass[preprint,journal]{vgtc}       % preprint (journal style)
\ieeedoi{10.1109/TVCG.2019.2934208}
\ifpdf%                                % if we use pdflatex
  \pdfoutput=1\relax                   % create PDFs from pdfLaTeX
  \usepackage{graphicx}                % allow us to embed graphics files
  \DeclareGraphicsExtensions{.pdf,.png,.jpg,.jpeg} % for pdflatex we expect .pdf, .png, or .jpg files
\else%                                 % else we use pure latex
  \usepackage{graphicx}                % allow us to embed graphics files
  \DeclareGraphicsExtensions{.eps}     % for pure latex we expect eps files
\fi%

\graphicspath{{figures/}{pictures/}{images/}{./}} % where to search for the images

\usepackage{microtype}                 % use micro-typography (slightly more compact, better to read)
\PassOptionsToPackage{warn}{textcomp}  % to address font issues with \textrightarrow
\usepackage{textcomp}                  % use better special symbols
\usepackage{mathptmx}                  % use matching math font
\usepackage{times}                     % we use Times as the main font
\usepackage{cite}                      % needed to automatically sort the references
\usepackage{tabu}                      % only used for the table example
\usepackage{booktabs}                  % only used for the table example
\usepackage{diagbox} % 加载宏包
\usepackage{amsmath}
\usepackage{color}
\onlineid{1065}

\vgtccategory{Research}
\vgtcpapertype{Empirical Study}

\title{Evaluating Perceptual Bias During Geometric Scaling of Scatterplots}

\author{Yating~Wei,
        Honghui~Mei,
        Ying~Zhao,
        Shuyue~Zhou,
        Bingru~Lin,
        Haojing~Jiang,
        and~Wei Chen}
\authorfooter{
\item Y. Wei, H. Mei, S. Zhou, B. Lin, and W. Chen are with The State Key Lab of CAD \& CG, Zhejiang University, Hangzhou, Zhejiang 310058, China. E-mail: \{weiyating, meihonghui, zhoushuyue, linbingru\} @zju.edu.cn, chenwei@cad.zju.edu.cn.
\item Y. Zhao, and H. Jiang are with School of Computer Science and Engineering, Central South University, Changsha, Hunan 410083, China. E-mail: \{zhaoying, gallows\}@csu.edu.cn
\item Corresponding authors: Wei Chen and Ying Zhao
}

\abstract{Scatterplots are frequently scaled to fit display areas in multi-view and multi-device data analysis environments. A common method used for scaling is to enlarge or shrink the entire scatterplot together with the inside points synchronously and proportionally. This process is called geometric scaling. However, geometric scaling of scatterplots may cause a perceptual bias, that is, the perceived and physical values of visual features may be dissociated with respect to geometric scaling. For example, if a scatterplot is projected from a laptop to a large projector screen, then observers may feel that the scatterplot shown on the projector has fewer points than that viewed on the laptop. This paper presents an evaluation study on the perceptual bias of visual features in scatterplots caused by geometric scaling. The study focuses on three fundamental visual features (i.e., numerosity, correlation, and cluster separation) and three hypotheses that are formulated on the basis of our experience. We carefully design three controlled experiments by using well-prepared synthetic data and recruit participants to complete the experiments on the basis of their subjective experience. With a detailed analysis of the experimental results, we obtain a set of instructive findings. First, geometric scaling causes a bias that has a linear relationship with the scale ratio. Second,
no significant difference exists between the biases measured from normally and uniformly distributed scatterplots. Third, changing the point radius can correct the bias to a certain extent. These findings can be used to inspire the design decisions of scatterplots in various scenarios.}%
\keywords{Evaluation, scatterplot, geometric scaling, bias, perceptual consistency}

\CCScatlist{ % not used in journal version
 \CCScat{K.6.1}{Management of Computing and Information Systems}%
{Project and People Management}{Life Cycle};
 \CCScat{K.7.m}{The Computing Profession}{Miscellaneous}{Ethics}
}

\renewcommand{\manuscriptnotetxt}{Manuscript received xx xxx. 201x; accepted xx xxx. 201x. Date of Publication
xx xxx. 201x; date of current version xx xxx. 201x. 
For information on obtaining reprints of this article, please send e-mail to: reprints@ieee.org.
Digital Object Identifier: xx.xxxx/TVCG.201x.xxxxxxx}

\vgtcinsertpkg

\begin{document}


\firstsection{Introduction}

\maketitle

\begin{figure}[ht]
\setlength{\abovecaptionskip}{1pt}
\setlength{\belowcaptionskip}{-5pt}
\vspace{-2mm}
\centering

\includegraphics[width = \linewidth]{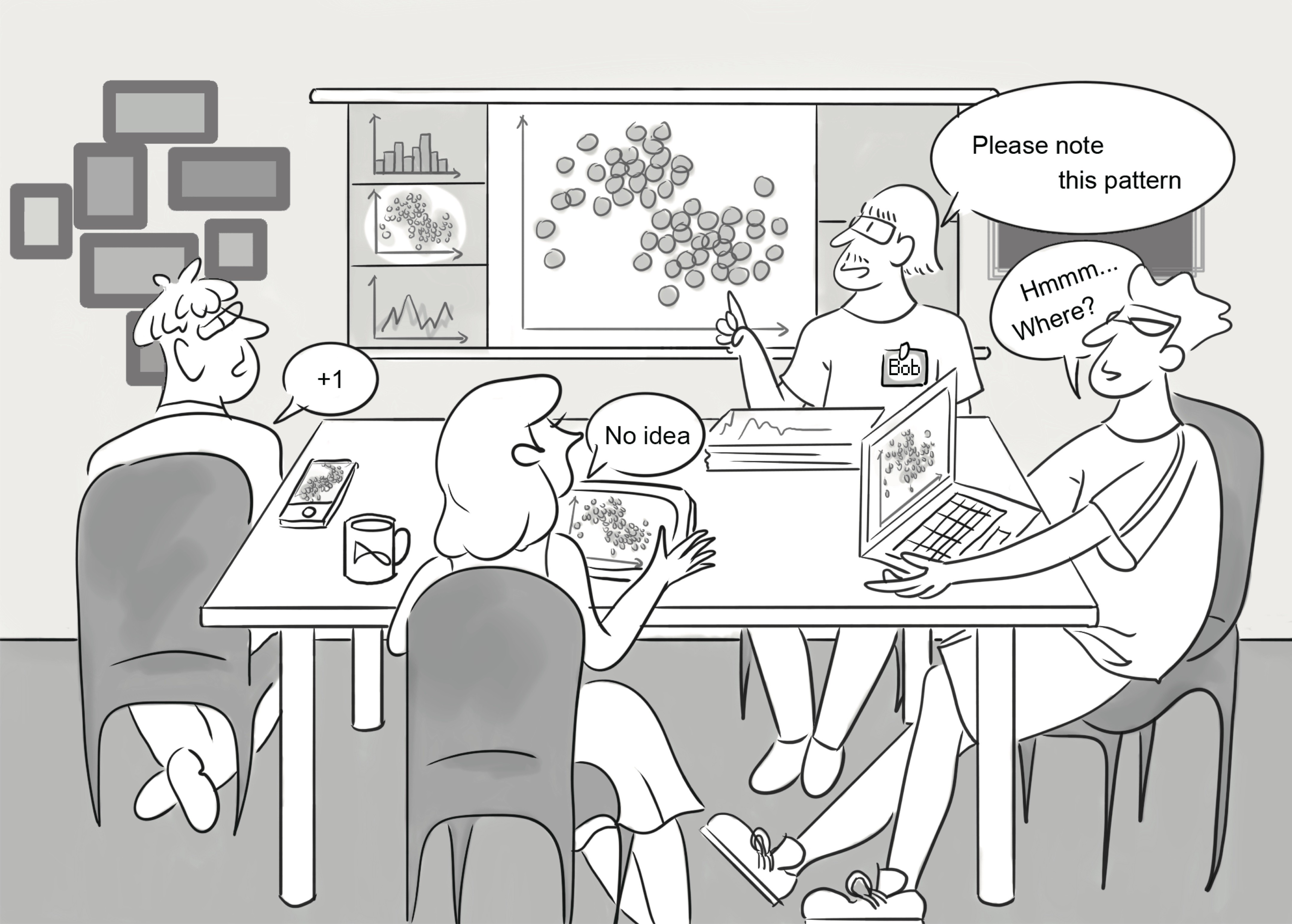}

\caption{
Scenario of scatterplot scaling. Bob finds a pattern of interest in a small scatterplot view and expands the scatterplot to a large view for further analysis. To exchange findings, he shares the scatterplot, and other collaborators examine the scatterplot using their own devices. The potential perceptual bias of scatterplots in various displays caused by scaling may lead to understanding inconsistency.
}
\label{fig1}

\end{figure}

%% \section{Introduction} %for journal use above \firstsection{..} instead
Scatterplots are a common type of visualization that supports the presentation and exploration of multi-dimensional data plotted on a two-dimensional (2D) plane~\cite{sarikaya2018scatterplots}. At present, various display devices~\cite{ni2006survey,chittaro2006visualizing,blascheck2019glanceable} are increasingly used in modern data analysis scenarios, and thus scatterplots are frequently scaled to fit different displays. In using multi-viewed business intelligence (BI) or visual analytics (VA) systems, analysts often expand a scatterplot of interest in thumbnail views to the main view to conduct in-depth exploration. Likewise, sharing a scatterplot across various displays of different sizes has nearly become a routine operation for communicating findings in collaborative data analysis~\cite{ardito2015interaction,chung2013comparison} (\autoref{fig1}).

The most straightforward method for scaling a scatterplot is to enlarge or shrink the entire plot together with the objects inside it synchronously and proportionally. This process is called geometric scaling. However, is geometric scaling always effective? Considering our practical experience and relevant research, we assume that the answer is no. Geometric scaling may cause a bias in visual perception, which is supposed to affect perceptual consistencies of data characteristics (e.g., numerosity, correlation, and clusters); therefore, this is detrimental to interactive data exploration. In Figure 1, two clusters are distinguished in Bob's scatterplot. However, they are indistinguishable when shown on mobile devices. In Figure 2(a), two scatterplots have the same amount of points, but people may feel that the right one has more points than the other. Moreover, existing research in psychology has demonstrated that a patch of points is recognized as sparse when the patch size increases~\cite{tibber2012number}, which supports our work.

Extensive studies have devoted to the design decisions of scatterplots (e.g., point color~\cite{wang2019optimizing,szafir2018modeling}, aspect ratio~\cite{cleveland1988shape,heer2006multi}, and animation~\cite{schulz2013design,brehmer2013multi}) in terms of empirical evaluations~\cite{rensink2010perception,tory2007spatialization,sedlmair2015data} or quantitative metrics~\cite{li2010model,sips2009selecting,bertini2011quality} to achieve improved information presentation in various scenarios. Moreover, the usage issues of scatterplots in mobile devices~\cite{chittaro2006visualizing}, high-resolution displays~\cite{ni2006survey}, and immersive environments~\cite{butscher2018clusters} have been extensively studied. However, scatterplot scaling, which is considered a special scenario for the usage and design decisions of scatterplots, has received minimal attention. To the best of our knowledge, no investigation has been systematically conducted on the potential perceptual bias during scatterplot scaling.

\begin{figure}[ht]
\setlength{\abovecaptionskip}{0pt}
\setlength{\belowcaptionskip}{-10pt}
\vspace{-2mm}
\centering

\includegraphics[width = 0.95\linewidth]{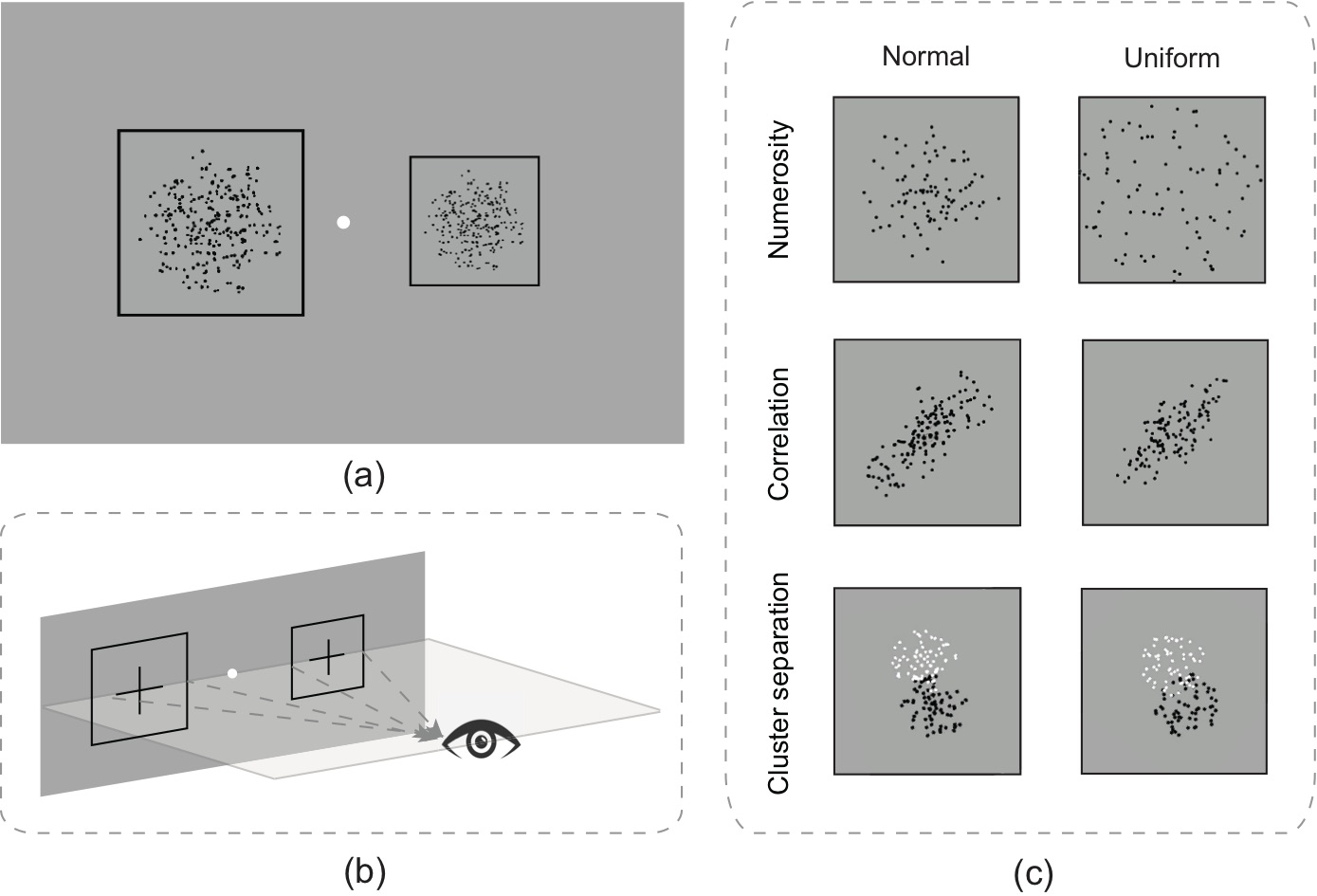}

\caption{
Stimuli and visual features. (a) Stimulus-pair consists of two scatterplots. In this example, the left figure is the original scatterplot and the right one is scaled by 63\%. (b) Positioning of the participant and stimuli. (c) Examples of synthetic scatterplots with three visual features and two distributions.
}
\label{fig2}
\end{figure}

To address this research gap, we conducted controlled experiments to study such a bias. Specifically, the bias refers to the deviation between the perceived and physical values of visual features. We emphasized on three visual features (i.e., numerosity, correlation, and cluster separation, as shown in \autoref{fig2}(c)) and proposed three hypotheses: geometric scaling may cause a bias; the bias can be affected by data distribution; and changing the radius of points may reduce the bias.

We adopted a two-interval forced choice (2IFC) method with a two-way staircase (2WS) design to simulate scatterplot scaling scenarios and collect participants' choice patterns of comparing a series of original-and-scaled scatterplot-pairs based on their subjective experience. Seven levels of scale ratio and point radius were selected, and scatterplots were generated using well-prepared synthetic data with 13 levels of visual feature, 2 distributions (normal and uniform). After several pilot studies to determine the experiments, we recruited 20 participants and conducted 3 rounds of experiments on each visual feature. Each round consisted of preparation, introduction and tutorial, pre-experiment, formal experiment, and subjective questionnaire.

We recorded the participants' choice patterns and subjective questionnaire answers as experimental results. We conducted a point of subjective equality (PSE) analysis to derive the quantitative biases from these choice patterns and carried out a series of statistical analyses on these derived biases. The analysis results showed that the first hypothesis was fully confirmed, the second hypothesis was fully negated, and the third hypothesis was partially confirmed. We also obtained other interesting findings from subjective questionnaires. We summarized these outputs systematically and discussed the limitations of this work and future directions.

In summary, we present the first attempt to understand the perceptual bias of scatterplots caused by geometric scaling. We contribute a carefully designed evaluation and a series of instructive findings, which may bring new considerations to the design decisions of scatterplots in various creation, exploration, and sharing scenarios~\cite{liao2017cluster,zhou2019survey,zhou2018visual,wang2018shadow}.

\section{Related Work}
\maketitle
\subsection{Perceptions of Scatterplot}
\label{section2.1}
Scatterplots are popular charts with a long history~\cite{friendly2005early,chen2014visual}. Research has shown that scatterplots are more effective than many other diagrams, especially when data are 2D~\cite{li2010judging,sedlmair2013empirical,xu2017efficient,lv2018crowd}. The popularity and usability of scatterplots are due to their simplicity and capability to allow users to easily perceive different visual features~\cite{sarikaya2018scatterplots}, which can be divided into three levels based on cognitive complexity. Low-level perceptions are obtained from direct visual stimuli (e.g., position, color and density/numerosity), which are called preattentive processing~\cite{ware2012information,healey2012attention}. Mid-level perceptions are understandings gained from stimuli patterns, such as correlation~\cite{rensink2010perception,rensink2017nature} and cluster separation~\cite{sips2009selecting,tatu2009combining,xia2018exploring}. High-level perceptions, such as memorability~\cite{borkin2016beyond}, aesthetics~\cite{harrison2015infographic}, and engagement~\cite{saket2016beyond}, are user cognition built upon the understandings. In this work, we investigate the effect of scatterplot scaling on perceived visual features. We examine one low-level (numerosity, which is perceived equally as density in most scenarios of visual analytics~\cite{anobile2014separate}) and two mid-level (correlation and cluster separation) visual features.

\subsection{Visual Encodings of Scatterplot}
Many works have studied design evaluation and automatic decision on the visual encodings of scatterplots. Design evaluation investigates the visual encodings of scatterplots in particular scenarios and provides empirical guidance on many facets, such as size and color of points~\cite{cleveland1985elements,ware2012information,szafir2018modeling}, aspect ratio~\cite{cleveland1988shape,heer2006multi,Wang2019Image}, selection of dimensions~\cite{tatu2010visual,bertini2011quality}, amenities~\cite{chan2010flow}, and interaction~\cite{saket2018evaluating,reach2019smooth} and animation~\cite{schulz2013design,brehmer2013multi}. Furthermore, many quality metrics have been developed for automatic design decision~\cite{behrisch2018quality, ma2018scatternet,mei2018design}. Quality metrics, such as those for clutter reduction~\cite{peng2004clutter} and cluster detection~\cite{tatu2010visual,kandogan2012just,xia2017ldsscanner}, measure the ability of diagrams to perform certain analytical tasks. However, existing studies have mainly focused on the design decisions when scatterplots are fixed and unchanged, whereas we are interested in the scenarios of changing display sizes.

The requirements for analyzing data on different displays, rather than only on desktop monitors, have increased with the development of display devices. Researchers have noted that display sizes strongly affect the usability of scatterplots~\cite{andrews2011information}. They have studied the effect of display sizes on design decisions of scatterplots, such as interactions on mobile phones~\cite{gutwin2004interacting,drucker2013touchviz}, cognitive processes on large high-resolution displays~\cite{andrews2010space}, and graph layouts in immersive environments~\cite{kwon2016study}. Moreover, the advances in display technology invoke new scenarios for exploratory analysis. On the one hand, large displays allow many diagrams to be presented simultaneously, thereby forming multi-view interfaces~\cite{yost2006perceptual,andrews2011information,reda2015effects}. On the other hand, given the application of spatially-aware and wireless connection techniques, multi-device combinations are increasingly used to provide collaborative data analysis involving multiple people. These new scenarios have led to frequent scatterplot scaling. However, few studies have investigated the effect of scatterplot scaling. In this work, we aim to find possible solutions to alleviate the effect of scatterplot scaling on visual consistency.

\subsection{Visual Biases of Scatterplot}
Perceptual bias is an important issue that has been extensively studied~\cite{alexander2017perceptual, ellis2018cognitive, zimmermann2016numerosity, valdez2018priming}; it considers the perceptual differences between situations, which include three cases. The first case refers to the differences between perceived visual features and quantitative statistics. Rensink et al.~\cite{rensink2010perception,rensink2017nature} found that the perceived correlation tends to underestimate the physical correlation; the correlation perception in scatterplots remarkably fits the Weber's law and is stable under the varying densities, aspect ratios, and distributions. Sedlmair et al.~\cite{sedlmair2012taxonomy} proposed that many factors may differentiate the perceived cluster separation from defined quality metrics. Alexander et al.~\cite{alexander2017perceptual} found that using  font size as a data encoding can influence the perception of underlying values. 
The second case is the inter-individual variations of perception. Many studies~\cite{gleicher2013perception,kay2016beyond,zhao2019evaluating} have proven that scatterplots have minimal inter-individual differences compared with other diagrams, such as radar graph~\cite{conati2008exploring,steichen2013user,shi2018novel} and parallel coordinates~\cite{kay2016beyond}. 
The third case refers to the differences between the perceived visual features of plots with various designs. Cleveland et al.~\cite{cleveland1982variables} found that the correlation perception can be enhanced by expanding the blank margins of scatterplots. Sophian and Chu proposed that the perceived numerosity decreases when dots cluster together~\cite{sophian2008people}. Valdez et al.~\cite{valdez2018priming} proved that priming and anchoring effects exist in perception tasks of scatterplots under certain circumstances. Yang et al.~\cite{yang2019correlation} indicated that people tend to use a small number of visual features when judging the correlation in scatterplots.
To the best of our knowledge, no study has systematically evaluated the perceptual bias caused by scatterplot scaling.

\section{Problems and Hypotheses}

A scatterplot often needs to be scaled to fit displays of different sizes in multi-view and multi-device data analysis environments. To maintain perceptual consistency among displays, the most straightforward method is to use geometric scaling, which either enlarges or shrinks the entire original scatterplot together with visual forms inside it proportionally. However, it is unclear that whether geometric scaling is effective in preserving perceptual consistency. Our basic observation is that geometric scaling will cause a perceptual bias. Such a bias can affect perceptual inconsistency in interactive data exploration.

According to Jerit~\cite{jerit2012partisan}, \emph{``Perceptual bias occurs when factual beliefs deviate from reality.''} Empirically, people frequently mismatch the perceived and actual values of visual features when a scatterplot is scaled. For example, if a scatterplot on a laptop is projected onto a curtain, then people may feel that the enlarged scatterplot has fewer points than the original one. Similarly, if a scatterplot is transferred from a desktop computer to a phone, then the clusters on the original scatterplot occasionally become blurred and unclear on the phone. Therefore, we suspect that geometric scaling can lead to a perceptual bias. Moreover, we are interested in investigating the factors that are affected by the bias. Is a certain law involved? Can we find a way to reduce the bias? In this work, we conduct controlled experiments to explore these issues. To guide our experimental design, we formulate the following hypotheses:

\textbf{H1:} We assume that geometric scaling causes a bias in the perceived visual features of scatterplots, including numerosity, correlation, and cluster separation. We also believe that the bias has a linear relationship with scale ratio ($Size_{scaled-one}$/$Size_{original-one}$). Thus, we seek to verify the existence of the bias and its inherent law related to the scale ratio. The three visual features are the most fundamental and important data characteristics that scatterplots allow users to perceive, and they have been thoroughly investigated by numerous studies (\autoref{section2.1}). The assumed linear relationship is based on our practical experience that the bias will be superior as the scale ratio increases or decreases.

\textbf{H2:} We assume that the bias caused by geometric scaling can be affected by data distribution. This hypothesis is used to explore whether the bias is related to data characteristics. We assume that scatterplots with various data patterns may cause different degrees of the bias. We select data distribution as our research object because it is a fundamental data characteristic. Specifically, we select normal and uniform distributions to perform a comparative analysis. Normal distribution that grows dense toward the center is treated as a different data pattern with uniform distribution.

\textbf{H3:} We assume that changing the radius of points can reduce the bias. We use this hypothesis to explore whether the bias can be reduced by changing visual encodings. The point radius, which is an important visual channel, is scaled proportionally in geometric scaling. We suppose that changing the radius in another way can reduce the bias.
\section{Experimental Design}
To test these hypotheses, we carefully designed three experiments. The first experiment (E1) measured the perceptual bias on the three visual features (H1). The second experiment (E2) examined whether the bias is affected by data distribution (H2). The third experiment (E3) aimed to find the effect of changes in point radius on the bias (H3). In this section, we introduce the experimental design in detail.

\subsection{Stimuli}
The stimuli were 2D scatterplots that shared some visual encodings and rendering options. The aspect ratio of all scatterplots was set to 1 to eliminate the interference caused by aspect ratio. All scatterplots were rendered with gray backgrounds and black thin borders~\cite{anobile2015mechanisms}, as illustrated in \autoref{fig2}(a). The domains of $x$ and $y$ dimensions in all scatterplots were normalized to [0,1]. Points in the scatterplots were represented by black circles. One exception was when testing cluster separation, in which the circles of two colors presented two clusters. The radii of circles in a scatterplot were the same. When a scatterplot was scaled, the visual encodings, including the width and height of the scatterplot, the length of axis, and the radii of points, were scaled proportionally. Our experiments used seven scale ratios, namely, 25\%, 40\%, 63\%, 100\%, 159\%, 252\%, and 400\%, which formed a geometric sequence for simulating the geometrical size changes of scatterplots. The scale ratio of the original scatterplot was maintained at 100\%. Other scatterplots were scaled by different scale ratios, of which three were enlarging and three were shrinking. We selected these scale ratios because the scale ratios of scatterplots can cover most scenarios considering common displays, including 4$-$6 $in$ mobile phones, 7$-$12 $in$ tablets, and 25$-$32 $in$ desktop monitors. To simplify the study, we used a unified desktop monitor to render the scaled scatterplots.

\subsection{2IFC \& 2WS Method}
We used a two-interval forced choice method with a two-way staircase design (2IFC \& 2WS) to simulate scatterplot scaling scenarios and measure people's subjective experience. 

\begin{figure}[ht]
\setlength{\abovecaptionskip}{3pt}
\setlength{\belowcaptionskip}{-5pt}
% \vspace{-2mm}
\centering

\includegraphics[width = \linewidth]{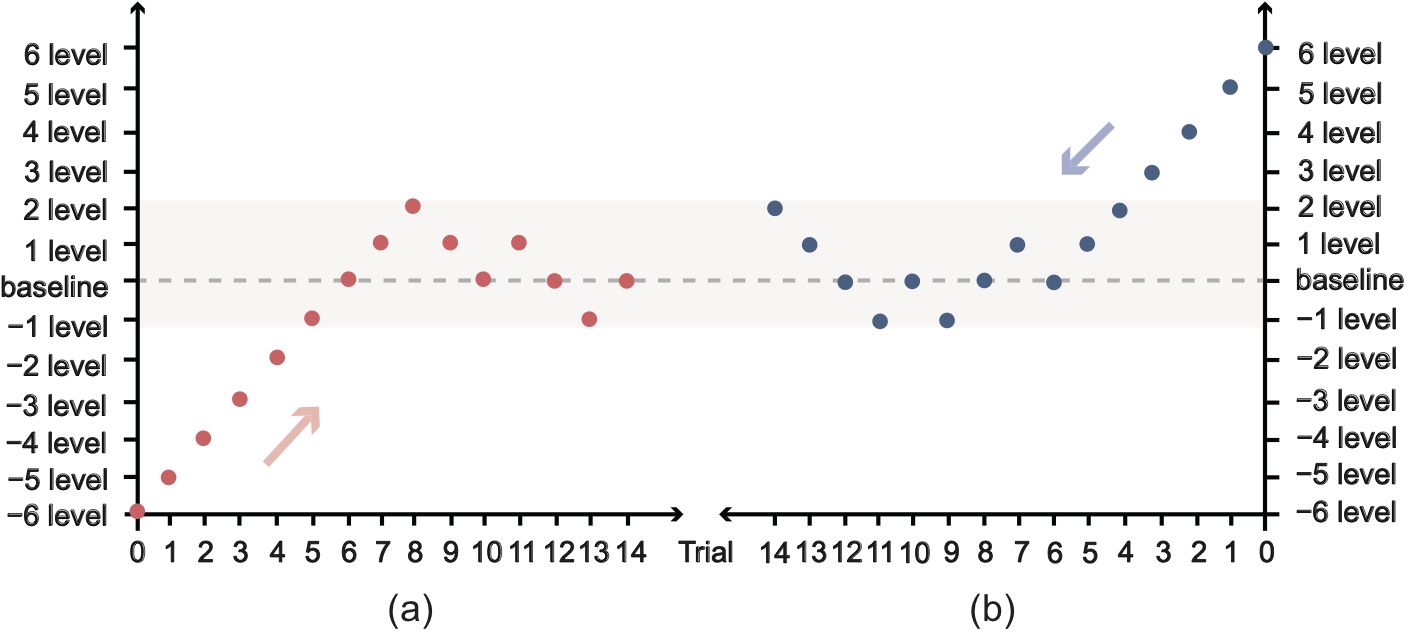}

\caption{
Example of a 2WS sequence. The $y$-axis shows visual feature levels. The $x$-axis shows the number of trials. (a) Fifteen trials of the forward-staircase. (b) Fifteen trials of the backward-staircase. The gray background shows the fluctuating interval of both staircases.
}
\label{fig3}
\end{figure}

The 2IFC is a method used to measure the subjective experience of a person through his/her choice pattern. In a 2IFC trial, the person is asked to judge between a reference and test stimulus-pair. The judgment is a subjective comparison to obtain the stimulus with a larger value of the visual feature. \autoref{fig2}(a) depicts an example. The left scatterplot with 256 points is the reference/original stimulus, and the right scatterplot with the same amount of points is the test stimulus that is scaled by 63\%. The person needs to select the one with a larger numerosity. Suppose the person selects the right one because he/she thinks that it has a larger numerosity, which deviates from the physical amount. Such a deviation may be caused by a perceptual bias. We can collect multiple judgments within a certain value range of the testing visual feature, and then measure the bias from the choice pattern. Therefore, a sequence of trials is required in a 2IFC task for each combination of visual feature and scale ratio. 

The construction of a reasonable 2IFC sequence involves two important considerations. (1) The levels of the testing visual feature must cover a suitable range. Too few levels make bias measurement less precise, and too many levels lead to an excessive number of trials. (2) The order of appearance of feature levels in a sequence of trials must be carefully designed. A random sequence requires many trials to obtain an accurate bias, whereas participants may find the regularity when levels are sorted.

Given that we had seven scale ratios, we used more levels than conventional settings (e.g., 7 in~\cite{tibber2012number} and 10 in~\cite{rensink2010perception}) to cover a large value range of visual features. We defined 13 feature levels for each visual feature, that is, \{$-$6 level, $-$5 level, ..., 0 level, ..., $+$6 level\}, 6 below and 6 above the baseline (0 level). In addition, referring to previous studies~\cite{tibber2012number,tibber2013sensitivity}, the levels of numerosity were defined through octave method~\cite{tibber2012number}, and the levels of the two other features were defined as an arithmetic series~\cite{rensink2017nature} (\autoref{section4.3}).

We used a 2WS design to determine the order of appearance of feature levels. A staircase design~\cite{garcia1998forced,ondov2019face} is that, the feature level of the test scatterplot varies to make the next trial difficult or simple, and the reference scatterplot is maintained at the baseline. The sequence can begin from either the $-$6 level (forward-staircase) or $+$6 level (backward-staircase); both terminate near the baseline, thereby forming a two-way staircase. \autoref{fig3} demonstrates the experimental result of a participant in a 2WS sequence. In the forward- or backward-staircase, the trials fluctuate within a certain interval around the baseline (the highlighted area) after several monotonic changes (the arrows) because the feature levels of the reference and test scatterplots are close to each other, thus making judgments difficult. Such a fluctuating choice pattern is the aim of our 2IFC \& 2WS method, through which the participant's subjective experience can be measured using PSE~\cite{tokita2010might,kim2014colour} to derive the bias (Section 5.1). In a word, the 2WS design provides a two-pass verification and eliminates the effect caused by the direction of level change. Moreover, the number of trials in each direction was set to 15 for a total of 30. This number of trials was greater than twice the number of feature levels, thereby guaranteeing the occurrence of fluctuation patterns.

In \autoref{fig2}(b), we positioned the center of the reference and test scatterplots symmetrically on the left and right parts of the center of the participant's viewpoint. Considering the displays of different sizes, we specified the width and height of a scatterplot in degrees of visual angle. The reference scatterplot extended $5^{\circ}$ of the visual angle vertically and horizontally. The scatterplot sizes of 7 scale ratios ranged from $1.25^{\circ}$ to $20^{\circ}$ of the visual angle. These sizes received comfortable feedback in our pilot studies and fit inside the near-peripheral vision~\cite{gutwin2017peripheral}.

\subsection{Experimental Variable and Data}
\label{section4.3}
We used synthetic data to generate scatterplots, as shown in \autoref{fig2}(c). The variables and data used in the three experiments were different.
\subsubsection{Experiment 1}
E1 was designed to measure the perceptual bias caused by geometric scaling on the three visual features of scatterplots: numerosity, correlation, and cluster separation. The variables of E1 included scale ratio, feature, and feature level. For each combination of scale ratio and feature, each participant was asked to finish a 2IFC \& 2WS sequence with 30 trials, thereby obtaining the choice pattern for bias measurement. Corresponding to 7 scale ratios, each participant completed 7 sequences for a total of 210 trials to detect the variation trend when the scale ratio changes. Three features resulted in 630 trials for each participant.

We generated a series of point sets for each feature. In each series, the corresponding visual feature was a variable with 13 levels of values to control the generation of 13 point sets, whereas other features remained unchanged. For each level, we generated a group of point set candidates and randomly selected one candidate for each trial to avoid participants' learning or fatigue effects. Specifically, the point sets for the three features were generated as follows.

\textbf{\emph{Numerosity:}} Point sets for numerosity were normally distributed point clouds with 13 numerosity levels. The baseline was 89 points. By extending the baseline numerosity in two ways, we obtained all 13 levels, which formed a 2-octave range~\cite{tibber2012number} as follows: 11, 15, 22, 31, 44, 63, 89, 127, 180, 256, 363, 515, and 730 points. The points were sampled from a 2D normal distribution with
% 公式
\begin{displaymath}
\setlength{\abovedisplayskip}{3pt}
\setlength{\belowdisplayskip}{3pt}
\mu = 
\begin{bmatrix}
0.5\\0.5
\end{bmatrix},\ 
\Sigma = \begin{bmatrix}
0.25&0\\0&0.25
\end{bmatrix}
\end{displaymath}
The positions of all points were limited in a constraint circle with (0.5, 0.5) as the center and 0.5 as the radius. If any point exceeded this constraint circle, then this point was removed and regenerated. Thus, the point cloud was normally distributed in the center of the scatterplot within a moderate range and with minimal overlap. 

\textbf{\emph{Correlation:}} Point sets for correlation were also normally distributed point clouds but with 13 levels of correlation coefficients, which formed an arithmetic series from 0.15 to 0.9 with a step of 0.0625 (baseline is 0.53). We generated point clouds for each correlation level by using a method similar to the data for numerosity, except that the number of points remained unchanged at 128, and the parameters of normal distribution were changed with
% 公式
\begin{displaymath}
\setlength{\abovedisplayskip}{3pt}
\setlength{\belowdisplayskip}{3pt}
\mu = 
\begin{bmatrix}
0.5\\0.5
\end{bmatrix},
\ \Sigma = \begin{bmatrix}
0.2&r\\r&0.2
\end{bmatrix}
\end{displaymath}
where $r$ is the target correlation coefficient. However, the correlation coefficient of generated point sets may float around target $r$. We only preserved those with correlation within
\begin{math}
r \pm 0.005
\end{math}.

\textbf{\emph{Cluster separation:}} For cluster separation, we considered the simple case of two clusters. Thus, we generated point sets that contained two normally distributed point clusters. We used a silhouette index $s$ to measure the cluster separation because it only depends on the actual partition of points rather than the clustering algorithm. We selected 13 levels of silhouette indices that formed an arithmetic series ranging from 0.11 to 0.65 with a step of 0.045 (the baseline is 0.38). Each point set contained 128 points, which were divided into two clusters of 64 points. However, directly generating a point set for a given silhouette index was difficult. Initially, we randomly generated two constraint circles, which centered at $(x_1,y_1)$ and $(x_2, y_2)$ with a radius of 0.4, for the two clusters. $x_1$, $y_1$, $x_2$ and $y_2$ were random numbers to ensure that the two constraint circles were completely inside the scatterplot. Then, we sampled points from two 2D normal distributions with

% 公式
\begin{displaymath}
\setlength{\abovedisplayskip}{3pt}
\setlength{\belowdisplayskip}{3pt}
\mu_1 = 
\begin{bmatrix}
x_1\\y_1
\end{bmatrix},
\ \mu_2 = 
\begin{bmatrix}
x_2\\y_2
\end{bmatrix},
\ \Sigma_1 = \Sigma_2 = \begin{bmatrix}
0.2&0\\0&0.2
\end{bmatrix}
\end{displaymath}
Lastly, we computed all silhouette indices of generated point sets and preserved these sets with silhouette indices within
\begin{math}
s \pm 0.005
\end{math}. The points of the two clusters were rendered black and white on a gray background. Such color settings reduced the participants' visual burden and enabled them to focus on cluster separation. 

\subsubsection{Experiment 2}
Given that E2 focused on the effect of data distribution, it shared the same variable settings with E1, and its data generation was similar to E1 in most settings. The difference was that the point sets were sampled from uniform distributions. Note that, for correlation tests in E2, we constructed a constraint ellipse and then sampled from a uniform distribution within the ellipse. The constraint ellipse had a center point at (0.5, 0.5), and the major axis went along the diagonal direction from the lower left to the upper right of the scatterplot. The length of the major axis was set to 1, whereas the length of the minor axis was a random number. Thus, we generated several candidate point sets and computed their correlation coefficients. Subsequently, we selected the candidates that satisfied the value requirements of each level with an error less than 0.005. 

\subsubsection{Experiment 3}
E3 was designed to investigate the effect of changing the point radius on the perceived visual features. It investigated three variables: point radius, feature, and feature level. Moreover, we selected two scale ratios as constants: 63\% and 252\%, that is, the $-$1 level and $+$2 level of 7 scale ratios, to test enlarging and shrinking cases. For the enlarging case, we selected the moderate $+$2 level. The shrinking cases used the $-$1 level because identifying small scatterplots ($\leq-2$ level) was difficult for participants. The radius ranged from 1.16 to 3.46, increasing geometrically and resulting in 7 levels (1.16, 1.39, 1.67, 2, 2.4, 2.88, and 3.46). For each combination of radius and visual feature, each participant was asked to finish a 2IFC \& 2WS sequence with 30 trials including 13 feature levels, thereby obtaining the bias of a certain feature for a given radius. For each feature, each participant was asked to complete 7 sequences of different radii to detect the variation trend of the bias when radius changed. Three features and two scale ratios resulted in 1,260 trials for each participant. The data used in E3 were the same as those in E1. The only difference was that the points were rendered with various radii.

\subsection{Participants and Apparatus}
Given that our experimental tasks were time consuming and visually strenuous, we recruited 20 participants (8 females and 12 males; undergraduate and graduate students with experience in data analysis using scatterplots; aged 19$-$25 years, median age: 21; with normal or corrected-to-normal vision) rather than using crowdsourcing approaches such as Amazon Mechanical Turk. We can observe the mental state of the participants and control the experimental process. Moreover, we provided an independent, quiet laboratory with a minimal external interference. We asked the participants to sleep early for at least three days before the experiments so that they would be fully rested and have full physical strength. We also prepared coffee and snacks to help the participants relax during breaks. The participants who completed the study were compensated with \$6 per hour.

All experiments were conducted on a Dell 25 $in$ U2515H monitor whose screen size is 21.8 $in \times $12.3 $in$ and resolution is $2560 \times 1440$ pixels, with a 60 Hz refresh rate and 50 $cd/m^{2}$ mean luminance. The participants viewed the stimuli at 59 $cm$ for the visual angle of approximately 1.25 arcmin per pixel. A standard wireless mouse and wired keyboard were used.

\subsection{Procedure}
\subsubsection{Pilot Study}
We invited 5 participants to conduct pilot studies for 3 times. These studies helped determine the details of the experiment design. In the pilot studies, we found that the stability of 2IFC judgments was much higher during the day than at night. This observation indicated that the human spirit and experimental environment had a great impact on the formulation of 2IFC judgments. Therefore, we arranged all experiments during the day.

To determine the experimental order, the first pilot followed the order of E1, E2, and E3. Each experiment contained all three features. We noticed that the participants had to refocus between features in a single experiment. Therefore, we used a feature-major order in the second pilot. Only one feature was tested in one round of E1$-$E3 to maintain the thinking-pattern continuity of the participants.

We originally planned to use a two-alternative forced choices (2AFC) method~\cite{tibber2012number} to display a scatterplot-pair (i.e., the left and right scatterplots appear concurrently), but we finally selected the 2IFC method (i.e., the left and right scatterplots appear sequentially) because the method enabled the participants to focus on comparing the scatterplot-pair for making a judgment, without worrying whether they could view two scatterplots simultaneously. 

Moreover, we repeatedly adjusted the data generation parameters (e.g., the value ranges and intervals of features), the interface and the procedure (e.g., adding practice trials and pre-experiment phase and adjusting the time span of breaks) through the pilot studies.

\subsubsection{Formal Study}
We designed a three-round experimental process for the formal study. The core variable in the three rounds was visual feature. That is, each participant performed the three experiments (E1, E2, and E3) of one feature in one round and completed all features (numerosity, correlation, and cluster separation) by performing three rounds. Four experimental phases were included in each round.

\textbf{Preparation:} An experiment instructor first led a participant into the laboratory room, and then assisted the participant in adjusting the position and height of the monitor to ensure that the point in the center of the screen was at eye level, as displayed in \autoref{fig2}(b).

\textbf{Introduction and tutorial:} This phase familiarized the participant with the experimental procedures, tasks, visual features, and data. This phase generally took approximately 10 min and comprised four steps.

\emph{STEP 1.} The instructor provided a brief description of the research purposes, related concepts, experimental procedure, and tasks.

\begin{figure}[ht]
\setlength{\abovecaptionskip}{0pt}
\setlength{\belowcaptionskip}{-8pt}
% \vspace{-3mm}
\centering

\includegraphics[width = \linewidth]{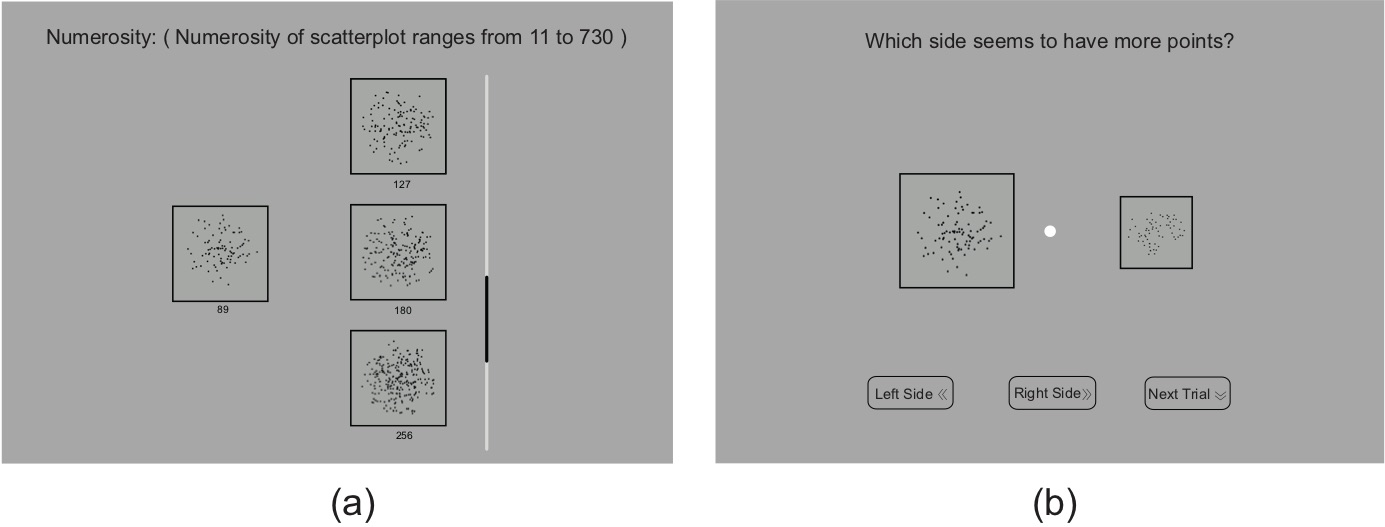}

\caption{
Interface of the experiment system. (a) Interface of the tutorial. In this example, the visual feature is numerosity. The scatterplot on the left is the reference. The 13 scatterplots on the right show 13 numerosity levels in ascending order, which can be scrolled through to browse. (b) Interface of the formal experiment. The judgment question is shown on the top. The scatterplot-pair and a center fixation point are located in the middle. Buttons for reporting judgments and proceeding to the next trial are found at the bottom. 
}
\label{fig4}
\end{figure}

\emph{STEP 2.} The instructor selected a visual feature from three features following Latin squares. Then, the instructor showed the prepared data samples of the selected feature and asked the participant to observe them for 2$-$3 min, as presented in \autoref{fig4}(a). Thus, the participant gained an intuitive presentation of the feature.

\emph{STEP 3.} The instructor explained the interactions of the system interface to the participant. The system interface is illustrated in \autoref{fig4}(b). Based on the 2IFC method, the left scatterplot was first presented for 250 ms and then disappeared. After displaying a center fixation point for 500 ms blank time, the right scatterplot was presented for 250 ms and then disappeared. Subsequently, the participant judged on the basis of his/her subjective experience. For numerosity, correlation, and cluster separation, the questions were ``Which side seems to have more points?'', ``Which side looks more relevant?'', and ``Which side seems to be more obvious in cluster separation?'', respectively. The interface supported two types of interactions to report the judgment: pressing the ``left/right arrow'' keys on the keyboard or clicking the ``Left/Right Side'' buttons on the screen using a mouse. The participant could modify the choice until he/she presses the ``down arrow'' key or click the ``Next Trial'' button to proceed to the next trial.

\emph{STEP 4.} The instructor provided 10 practice trials of the current visual feature and asked the participant to complete the trials to be familiarized with the interactions and rhythm of making judgments. 

\textbf{Pre-experiment:} After finishing the tutorial, the participant was required to perform a short pre-experiment for the current feature to test whether the participant had been fully prepared. The pre-experiment contained 2 (levels) $\times$ 30 (trials of a sequence) $\times$ 4 (E1, E2, E3 [scale ratio = 63\%], and E3 [scale ratio = 252\%]) trials. After completing the pre-experiment, the instructor analyzed the result immediately. If the result did not reach the standard (\autoref{section5.1}), then the experiment was suspended. In general, a poor pre-experiment result was due to poor focus or imperfect task understanding. If the participant expressed fatigue, then he/she would be asked to rest for at least 2$-$3 hours. If the participant was confused with tasks, then back to the previous phase. This phase took approximately 6 min.

\textbf{Formal experiment:} The participant clicked the ``Start'' button to begin making judgments individually. After completing all trials of one experiment, the participant rested for 2$-$3 min and then proceeded to the next experiment. The order of three experiments (E1$-$E3) was fully counterbalanced across the participants. Each participant was required to complete 840 trials in one round (7 (levels) $\times$ 30 (trials of a sequence) $\times$ 4 (E1, E2, E3 [scale ratio = 63\%], and E3 [scale ratio = 252\%])). A full round of formal experiments, including breaks, took approximately 90 min to complete.

After one round, the participant was required to fill out the subjective questionnaire of the current round. A short interview was conducted subsequently to obtain additional subjective information. Then, the participant was given 1 h of rest. He/she was allowed to leave the laboratory room to take coffee and snacks to relax and relieve fatigue. Afterward, the participant proceeded to the next round. After three rounds, a formal interview was conducted. All three rounds had 2,520 trials (840 (trials of a round) $\times$ 3 (features)), and generally took approximately 6.5 h. Moreover, 7 scale ratios/point radii were fully counterbalanced using Latin squares among the participants, and the positions of reference and test scatterplots were completely random.

\subsubsection{Subjective Questionnaire}
The participants were asked to fill out a questionnaire to collect their subjective feelings on the current examined visual feature after completing a round of experiments. The questionnaire was set with two types of questions. (1) The first type inquired about the subjective feeling of the overall difficulty level (DL) of making 2IFC judgments. For example, ``Do the changes of scatterplot size make your judgments on the perception of normal-distributed numerosity difficult? If so, then how difficult is it?'' The participants rated it using a five-point Likert scale ranging from 1 (the lowest difficulty) to 5 (the highest difficulty). (2) The second type inquired about the subjective experience of the difficulty tendency (DT) of making 2IFC judgments with different scale ratios or point radii. For example, ``How does the difficulty of making judgments change on the perception of normal-distributed numerosity with the changes in scatterplot size from small to large?'' The participants were given five options: A. increasing, B. decreasing, C. increasing after decreasing first, D. decreasing after increasing first, and E. not sure. In the questionnaire, four specific questions were used for each of the two question types with respect to E1, E2, E3 (scale ratio = 63\%), and E3 (scale ratio = 252\%). The four specific questions were slightly different in experimental variables. Specifically, we asked about the normal distribution for E1, uniform distribution for E2, and point radius for E3. In summary, each participant was asked to solve 8 subjective questions for a visual feature and 24 questions for the three visual features.

\section{Experimental Results}

\subsection{Analysis Approach}
\label{section5.1}
We collect all objective and subjective results from the experiments. Objective results record the information of the participants who complete each 2IFC trial, including choice, location of reference scatterplot, experimental variables, and completion time. Subjective results are extracted from the questionnaires, including 24 questions and interview records for each participant. We analyze the results in the following three main aspects.

\begin{figure}[ht]

\centering
\setlength{\abovecaptionskip}{3pt}
\setlength{\belowcaptionskip}{-5pt}
\includegraphics[width=\linewidth]{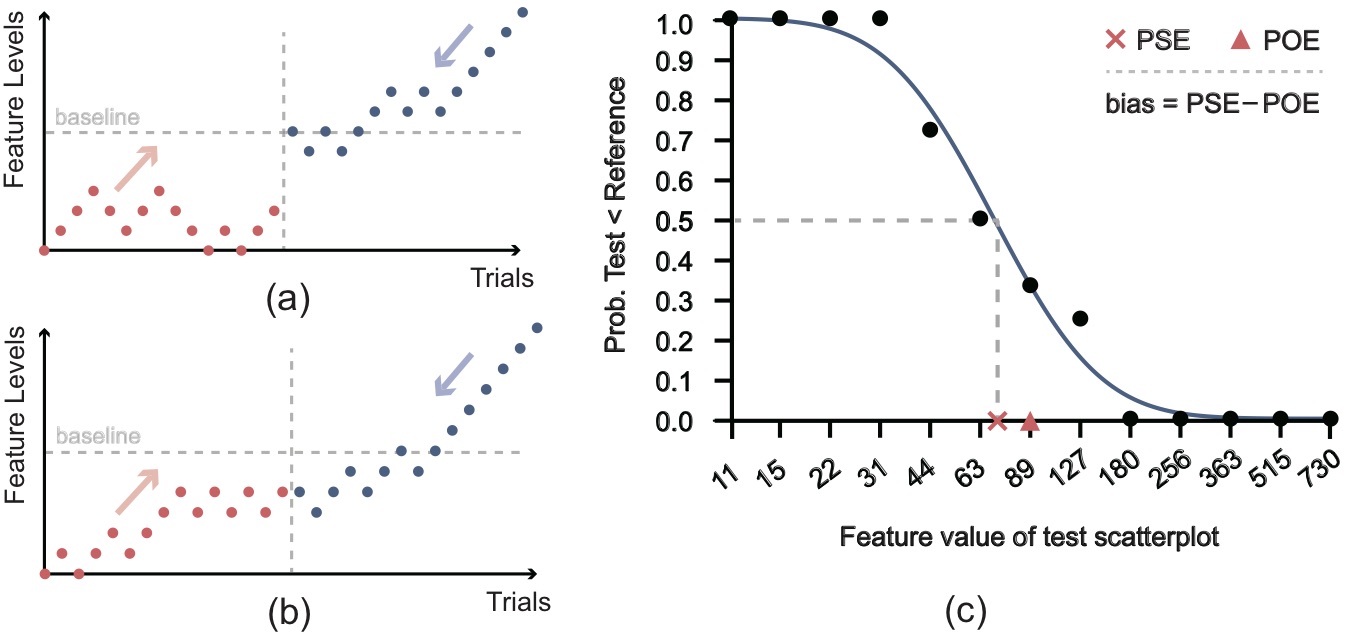}

\caption{
Examples of abnormal choice patterns: (a) fiercely fluctuating and (b) early fluctuating. (c) Illustration of the bias measurement. The plot is drawn on the basis of the choices of a sequence of trials made by a certain participant on numerosity at scale ratio = 40\%, where the $x$-axis is the actual feature value of the test scatterplot, and the $y$-axis is the probability of selecting ``the feature value of test scatterplot is perceived to be smaller than the feature value of the reference scatterplot.'' Furthermore, a data point represents the proportion of ``test $<$ reference'' choices made by the participant at a feature value of the test scatterplot, and the blue curve is the best-fitting cumulative Gaussian function of all data points. Thus, the POE is the baseline feature value point on the $x$-axis, whereas the PSE is the point on the best-fitting curve where $y$ = 0.5. The bias is the value of the PSE minus the POE.}
\label{fig5}
\end{figure}

\textbf{Abnormal choice patterns identification.} Some participants could be distracted occasionally within hours of 2IFC judgments, resulting in abnormal choice patterns in a sequence of trials. Two representative examples are depicted in \autoref{fig5}(a$-$b). Such abnormalities would have a serious effect on the accuracy of the subsequent bias measurement. We perform a variance analysis to identify fiercely fluctuating choice patterns. We compute the variance of the feature levels of 10
essential trials (the last five trials of forward- and backward-staircases)
of a sequence. A large variance ($>2$) typically indicates the occurrence
of fiercely fluctuating choice patterns on the basis of the feedback from the participants in the pilot studies. In addition, we manually examine a small amount of sequences with early fluctuating choice patterns by observing whether the sequence fluctuates from the second or third trial. Finally, we identify 16\% abnormal sequences. We mark these sequences as outliers and replace the biases measured from them with the mean of biases of all participants with the same experimental variables.  

\textbf{Bias measurement.} Based on the human psychometric function~\cite{dakin2011common} and Weber's law~\cite{kay2016beyond}, the bias of a certain participant on a visual feature at a scale ratio/point radius can be measured quantitatively from his/her choices of a sequence of trials using point of subjective equality (PSE) and point of objective equality (POE). In this case, when the test and reference scatterplots of a trial appear subjectively to have the same feature value, the participant makes a random choice. Thus, the PSE is the 0.5 probability point of selecting ``test $<$ reference'' in a sequence of trials. The POE is the baseline feature value in which the actual feature values of the test and reference scatterplots are objectively equal. Therefore, if scatterplot scaling causes a perceptual bias, then the PSE in a 2WS sequence is set apart from the POE with a certain interval. The interval is exactly the bias, which equals the PSE minus the POE. As shown in \autoref{fig5}(c), the POE is 89 and the PSE is 65.86. At the PSE, the participant perceived that the numerosity values of two scatterplots were equal, yielding a negative bias ($-$23.14 = 65.86 $-$ 89). However, the actual numerosity value (65.86) of the test scatterplot was less than that (89) of the reference one. This finding indicates that the shrunk test scatterplot (scale ratio = 40\%) was perceived with a relatively large value of numerosity.

\textbf{Significance analysis.} We conduct two significance analyses for each experiment (E1, E2, and E3 [ratio = 63\% and 252\%]) to examine the significance of the objective results. The first is for the biases among the participants. We first use the Shapiro$-$Wilk test to examine the normality and find that most results did not follow the normal distributions ($p < 0.05$). Then, we use non-parametric Friedman and ANOVA tests for examination. The results show that, for each experiment and visual feature, no significant differences are found in the biases of various participants. Thus, the biases of different participants can be treated equally. The second significance analysis is for the biases among different scale ratios/point radii. The Shapiro$-$Wilk test shows that all results do not follow the normal distributions ($p < 0.05$). Thus, we use non-parametric Friedman tests to examine whether significant differences occur among the biases of 7 scale ratios/point radii for each visual feature.

\begin{figure*}[!ht]

\centering
\setlength{\abovecaptionskip}{3pt}
\setlength{\belowcaptionskip}{-5pt}
% \vspace{-2mm}
\includegraphics[width = 0.95\linewidth]{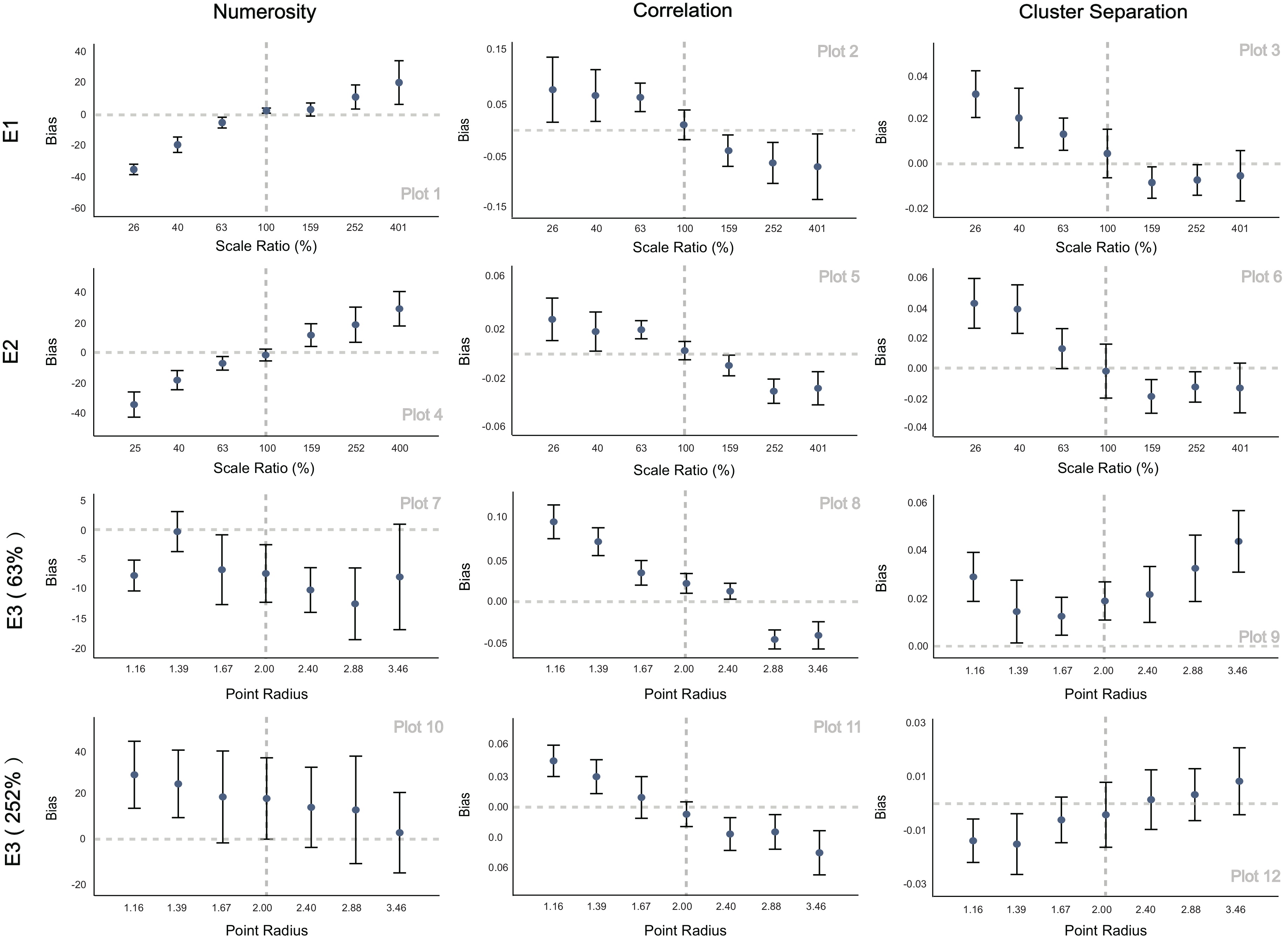}

\caption{
Objective results of the bias measurement with regard to experiments (E1, E2, E3 [scale ratio = 63\%], and E3 [scale ratio = 252\%]) and three visual features (numerosity, correlation, and cluster separation). A plot presents the biases of 7 scale ratios/point radii for a certain experiment and feature, in which the $x$-axis is a logarithmic (base 10) axis of scale ratio for E1 and E2 or point radius for E3 because 7 levels of scale ratios/point radii form a geometric sequence. The $y$-axis is the value of a bias. A point represents the mean of biases of all participants at a certain scale ratio/point radius. The error bar of a point encodes a 95\% confidence interval, and two dashed lines in a plot represent the baselines of 0 bias and 100\% ratio/2.0 radius.
}
\label{fig6}
\end{figure*}

\begin{figure*}[tb]
\setlength{\abovecaptionskip}{0pt}
\setlength{\belowcaptionskip}{-5pt}
% \vspace{-3mm}
\centering

\includegraphics[width = 0.95\linewidth]{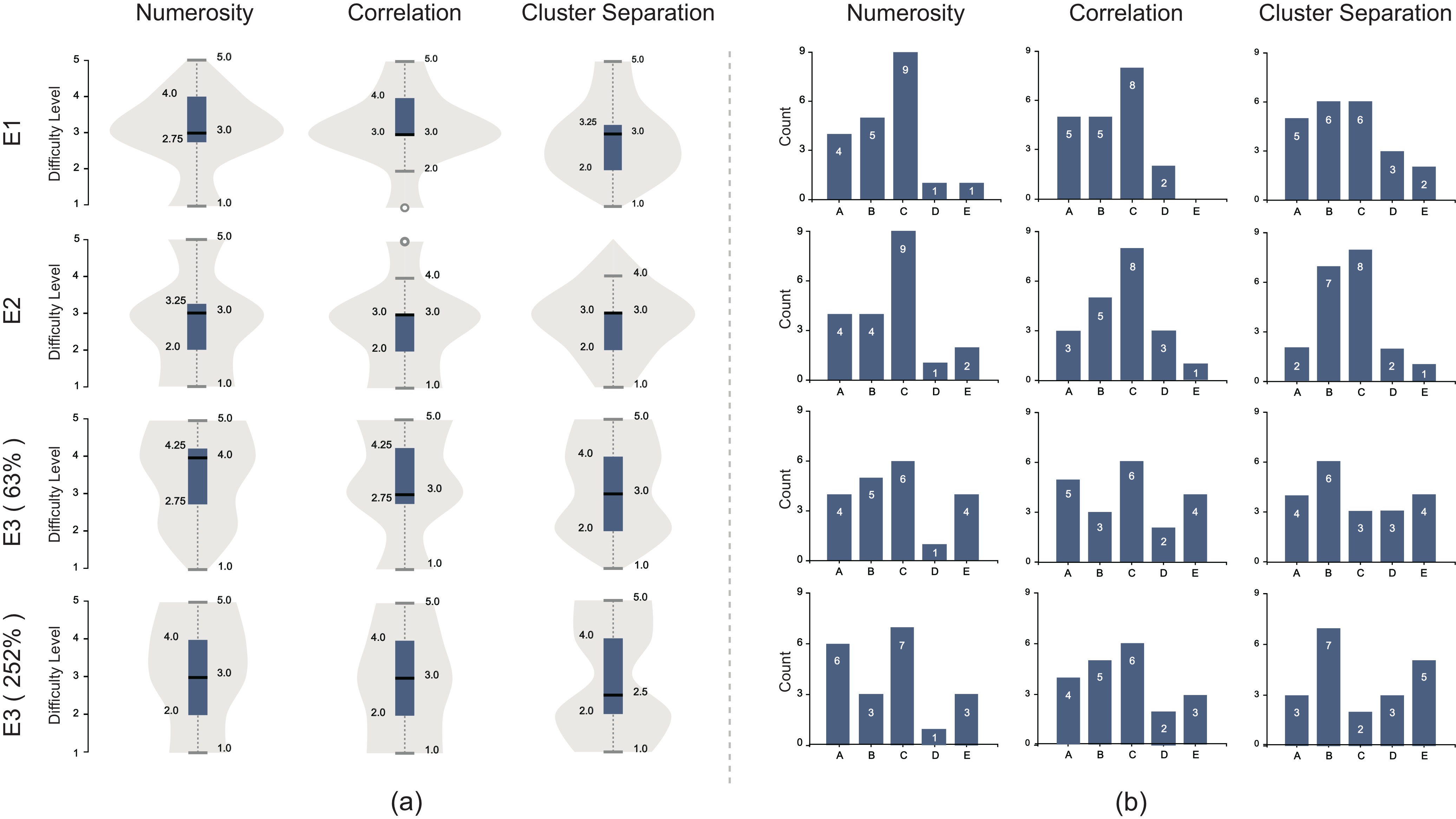}

\caption{
Subjective questionnaire results. (a) Results of DL questions. A violin plot, which consists of a box plot and an area plot, presents the results of all participant ratings on the DL questions for an experiment and feature. The $y$-axis indicates the judgment difficulty levels ranging from 1 (lowest difficulty) to 5 (highest difficulty). A box plot presents the median, upper quartile, lower quartile, 1.5 IQR of the lower quartile, and 1.5 IQR of the upper quartile of ratings. An area plot presents the kernel density distribution of ratings on the five difficulty levels. (b) Results of DT questions. A bar chart presents the results of all participants who select the tendency options of the DT question for an experiment and feature. The $x$-axis encodes the five options (A. increasing, B. decreasing, C. increasing after decreasing first, D. decreasing after increasing first, and E. not sure) and the $y$-axis encodes the counts of the participants selecting a certain option.
}
\label{fig7}
\end{figure*}

\subsection{Objective Result Analysis}
\label{section5.2}
In this section, we test against the three hypotheses with the results of the bias measurement (\autoref{fig6}) and significance analysis. We take Plot 1 in Figure 6 as an example to explain how to read the bias measurement results. The bias is close to 0 when the scale ratio is 100\%, indicating that the participants were nearly unbiased on the perception of numerosity when the test scatterplot was not scaled. The biases are negative and gradually move away from 0 bias when the scale ratio decreases (from 100\% to 63\%, 40\%, and 25\%), denoting that the participants perceived a higher numerosity than the physical numerosity in the shrunk test scatterplot and the difference between the physical and perceived numerosity increased gradually. The biases are positive and gradually move away from 0 bias when the scale ratio increases (from 100\% to 159\%, 252\%, and 400\%). This result signifies that the participants perceived a lower numerosity in the enlarged test scatterplot and the difference increased gradually.

\textbf{Hypothesis 1.} We assume that geometric scaling causes a bias in perceiving the three visual features, and the bias may have a linear relationship with the scale ratio. This hypothesis is fully confirmed.  

With the existence of the bias, as shown in Plots 1, 2, and 3 in \autoref{fig6}, all features show the biases to a certain extent when the scale ratio is not at 100\%, whereas the biases are very close to 0 when the scale ratio is at 100\%. This observation indicates that enlarging and shrinking the test scatterplot affect the perceptual consistency of the three features. 

In terms of the relationship between the bias and scale ratio, the results of significance analysis show significant differences in the biases of 7 scale ratios for all features (numerosity: $\chi ^2$ = 85.987, $p = 0.000 < 0.05$; correlation: $\chi ^2$ = 47.244, $p = 0.000 < 0.05$; cluster separation: $\chi ^2$ = 58.140, $p = 0.000 < 0.05$), indicating that a relationship must occur between the bias and scale ratio. We conduct a linear regression analysis on the biases of 7 scale ratios by feature. The results of fitting index $R^2$ (numerosity: $R^2 = 0.940 > 0.9$; correlation: $R^2 = 0.969 > 0.9$; cluster separation: $R^2 = 0.920 > 0.9$) denote that the bias has a significant linear relationship with the scale ratio for each feature.

We analyze further the variation trend of the bias by feature when the size of the test scatterplot gradually increases from the smallest scale ratio (25\%) to the largest (400\%) and obtain the following findings.

% \begin{enumerate}
% \renewcommand{\labelenumi}{(\theenumi)}
\textbf{\emph{Numerosity:}} The bias presents a positive growth trend with the increase in the scale ratio (Plot 1 in \autoref{fig6}). This result means that the perceived numerosity decreases while the test scatterplot is gradually enlarged. On the basis of the occupancy model~\cite{allik1991occupancy} and Steven's power law~\cite{stevens1986psychophysics}, the perceived numerosity in a scatterplot can be identified with the area occupied by points, and the perceived change in the point area is generally less significant than the perceived change in the scatterplot size when the scatterplot is scaled. That is, the perceived increase in the point area cannot keep pace with the perceived increase in the scatterplot size when the test scatterplot is enlarged, thereby leading to an underestimation of numerosity in the enlarged test scatterplot. The test scatterplot requires a high physical numerosity to be perceived as the same numerosity of the reference one.

\textbf{\emph{Correlation:}} The bias has a negative growth trend with the increase in the scale ratio (Plot 2 in \autoref{fig6}). This result means that the perceived correlation increases while the test scatterplot is gradually enlarged. The participants said that they perceived a correlation mainly by observing the area and major axis length of the ellipse formed by points~\cite{yang2019correlation}. According to Steven's power law~\cite{stevens1986psychophysics}, the perceived change in the ellipse area is less significant than the perceived change of the ellipse length when a scatterplot is scaled. That is, the perceived ellipse is relatively thin in the enlarged test scatterplot, and the test scatterplot requires a small physical correlation to be perceived as correlated as the reference one.

\textbf{\emph{Cluster separation:}} The bias has a negative growth trend with the increase in the scale ratio (Plot 3 in \autoref{fig6}). The participants expressed that they perceived cluster separation mainly by observing the number of points in the overlap area of two clusters. Similar to numerosity, the perceived change in the overlap area is less significant than the perceived change in the scatterplot size when a scatterplot is scaled. That is, two clusters are perceived with a relatively large level of cluster separation in the enlarged test scatterplot. Moreover, the negative growth trend of the bias becomes flat on the right side of the 100\% scale ratio. The participants likely encounter less difficulty in identifying the points in the overlap area when the size of the test scatterplot is sufficiently large. 

\textbf{Hypothesis 2.} We assume that the bias can be affected by data distribution. We conduct pairwise $t$-tests to examine the significant differences between the biases of E1 (normal distribution) and E2 (uniform distribution) by feature and scale ratio. Then, we apply a Bonferroni correction to reduce the significance level from $p = 0.05$ to 0.00714 (0.05$/$7) because we divide the results of the bias measurement of each feature by 7 scale ratios in the $t$-test. No significant differences ($p < 0.00714$) are found between the biases of E1 and E2 at 7 scale ratios for the three features. This hypothesis is fully negated. We think that the visual patterns formed by a normal distribution may have inefficient difference from those formed by a uniform distribution. 

\textbf{Hypothesis 3.} We assume that changing the point radius can reduce the bias. In E3, the test scatterplot was scaled at a fixed scale ratio, and its point radius changed in 7 levels, whereas the reference scatterplot had an unchanged baseline radius $(r = 2)$, which was the point radius for all scatterplots in E1. Two scale ratios (63\% and 252\%) were selected to test the enlarging and shrinking cases. This hypothesis is partially confirmed. 

For the bias at the baseline radius $(r = 2)$, we find a certain degree of bias in each of Plots 7$-$12 in \autoref{fig6}. This result is reasonable because the test scatterplot is consistently scaled in E3. The results of significance analysis show that significant differences exist in the biases of 7 point radii for each feature (for scale ratio = 63\%, numerosity: $\chi ^2$ = 19.330, $p = 0.004 < 0.05$, correlation: $\chi ^2$ = 89.688, $p = 0.000 < 0.05$, and cluster separation: $\chi ^2$ = 30.021, $p = 0.000 < 0.05$; for scale ratio = 252\%, numerosity: $\chi ^2$ = 37.104, $p = 0.000 < 0.05$, correlation: $\chi ^2$ = 53.314, $p = 0.000 < 0.05$, and cluster separation: $\chi ^2$ = 17.593, $p = 0.007 < 0.05$).

We then analyze whether changing the point radius can reduce the bias for the three features.

\textbf{\emph{Numerosity:}} For scale ratio = 63\%, when the point radius decreases from 2 to 1.39 step by step, the bias gradually approaches 0 from $-$6.23 to $-$0.0712, as exhibited in Plot 7, thereby indicating that the negative bias can be corrected by reducing the point radius when the test scatterplot is shrunk. This correction trend is consistent with the results of E1 that the perceived numerosity is relatively large when the test scatterplot is shrunk, and the radius must be reduced to correct the bias. Notably, the bias largely deviates from 0 suddenly when the point radius becomes 1.16, indicating that such a correction trend only exists in a certain range of point radii. For scale ratio = 252\%, when the point radius increases from 2 to 3.46 step by step, the bias gradually reduces from 16.02 to 3.204, as demonstrated in Plot 10, denoting that the positive bias can be corrected by increasing the point radius when the test scatterplot is enlarged. This correction trend is consistent with the results of E1 that the perceived numerosity is relatively small in the enlarged test scatterplot, and increasing the point radius can reduce the bias. In addition, the bias is not very close to 0 even at $r = 3.46$ likely because we do not sufficiently test the levels of point radius. In summary, changing the point radius within a certain range can slightly correct the perceptual bias of numerosity.

\textbf{\emph{Correlation:}} For scale ratio = 63\%, when the point radius increases from 2 to 2.4, the bias reduces from 0.021 to 0.0105, as illustrated in Plot 8. For scale ratio = 252\%, when the point radius decreases from 2 to 1.67, the bias changes from $-$0.00735 to 0.0047, which approaches 0, as demonstrated in Plot 11. These results indicate that changing the point radius can correct the perceptual bias of correlation. However, the correction trend of correlation only occurs within two adjacent levels of point radius, whereas such a trend of numerosity occurs within three adjacent levels. The reasons are that the biases of correlation at the baseline radius are highly close to 0, and the correlation perception may be insensitive to the change in point radius.

\textbf{\emph{Cluster separation:}} For scale ratio = 63\%, the bias reduces from 0.0186 to 0.0129 when the point radius decreases from 2 to 1.67, as shown in Plot 9, but the bias suddenly increases to 0.028 when the point radius continues to decrease to 1.16. For scale ratio = 252\%, the bias changes from $-$0.00418 to 0.0019 when the point radius increases from 2 to 2.4, as presented in Plot 12. However, as the point radius continues to increase, the bias also increases. These results indicate that the bias of cluster separation can be corrected by changing the point radius within a certain range. Similar to correlation, the correction effect of cluster separation only occurs within two adjacent feature levels.

\subsection{Subjective Result Analysis}
In this section, we analyze the results of the subjective questionnaires.

\textbf{Experiment 1.} For the DL questions, Level 3 (moderate difficulty) obtains the most ratings for all features. A comparison of the ratings among the three features shows that cluster separation obtains a relatively low overall rating. This condition reflects that geometric scaling has a stronger influence on the perception of numerosity and correlation than that of cluster separation. Many participants commented that, \emph{``In the trials of cluster separation, two clusters are marked with different colors, making the judgments easy.''} For the DT questions, Option C gains the most selections for numerosity and correlation; this finding is consistent with the bias trend analysis of H1 (\autoref{section5.2}). In terms of cluster separation, Options B and C obtain the same amount of selections. Both options are reasonable. In Plot 3 of \autoref{fig6}, the bias tends to decrease first and then increases with the scale ratio. However, the rate of increase decreases after the scale ratio reaches 100\%.

\textbf{Experiment 2.} The subjective results of E2 are similar to those of E1. With distributions, the overall judgment difficulty level is slightly lower in E2 than in E1. This result can be observed by comparing the box plots of the two experiments by feature in \autoref{fig7}(a). A participant reported, \emph{``The boundary of points of uniform distribution is clearer than that of normal distribution, which is beneficial to the judgments.''}

\textbf{Experiment 3.} For the DL questions, more participants rated Levels 4 and 5 for all features compared with E1 and E2. We speculate that the participants had difficulty focusing on various point radii and scaled scatterplots simultaneously in E3. This speculation can be verified by the results of the DT questions. The number of participants who selected Option E for all features was clearly more than that of E1 and E2. According to a participant, \emph{``I had difficulty knowing exactly whether the point radius becomes larger or smaller as the scatterplot itself has been scaled.''} Moreover, the overall difficulty level is higher in the 63\% scale ratio than in the 252\% one, as depicted in \autoref{fig7}(a). Many participants reported the ease of making judgments in a large-scale ratio. As a participant said, \emph{``The details are clear enough in an enlarged scatterplot, and I am more confident about my judgments.''}

\subsection{Summary}
The objective results reflect three main findings. First, geometric scaling causes a bias on the perception of the three visual features (numerosity, correlation, and cluster separation). The bias has a linear relationship with the scale ratio; that is, the absolute value of bias is enlarged when the scale ratio increases or decreases. Second, no significant difference occurs between the biases measured from normally and uniformly distributed scatterplots. Third, changing the point radius can correct the bias. Such a correction only appears
in a certain radius range, and the correction effect is large in the cases of numerosity and at a large scale ratio.

Our findings from the results of subjective questionnaires are also threefold. First, the participants were aware of the effect of geometric scaling on the perceived three features, of which numerosity was most affected. The participants could roughly realize the relationship between the bias and scale ratio. Second, several participants reported that the biases were slightly less affected by geometric scaling in the uniformly distributed scatterplots than in the normally distributed ones. Third, the participants were ambiguous about the bias correction effect of changing the point radius, but some participants were able to easily judge the enlarged scatterplots.

\section{Discussion}
In this section, we discuss the lessons learned from this study. We also summarize the limitations and suggest directions for further work.

\textbf{Hypothesis formulation.} We studied the low- and mid-level perceptions of scatterplot (H1), except high-level. We believe that the bias is ubiquitous during scatterplot scaling but can be presented in various forms. We preliminarily investigated the effect of normal and uniform distributions (H2). Diverse data distributions are worth exploring in the future. We only investigated the effect of point radii on the bias (H3). Other visual channels require further exploration, such as color, contrast~\cite{tibber2012number}, opacity~\cite{matejka2015dynamic}, and luminance~\cite{ross2010vision}.

\textbf{Experimental design.} The most important lesson learned from the experiment design was the importance of pilot studies. The 2AFC method, the random sequence of trials, and the experiment order from E1 to E3, were verified to be unreasonable in our pilot studies. We found that the performance of the participants would decrease over time if the experiments were conducted continuously. To minimize the effect of fatigue on the experimental results, we set up a variety of rigorous rest mechanisms. After completing all trials in one experiment, the participants were asked to rest for 2$-$3 min before proceeding to the next experiment. After a round of experiments, the participants were given an hour of rest, and a pre-experiment was conducted before the next round to test the participants' mental state. The participants were allowed to complete all experiments within a few days if they reported a strong feeling of fatigue or encountered unexpected situations during the experiments. Moreover, all scatterplots in the experiments were simplified to be shown on a unified desktop monitor. This practice benefited controllable experiments but reduced the ecological validity. We can further explore how different display devices and screen resolutions influence the bias caused by geometric scaling.

\textbf{Experimental result analysis.} The biases measured from a small number of abnormal sequences were replaced with the mean of biases of all participants with the same experimental variables. We carefully examined these sequences to ensure all of them have obviously abnormal choice patterns, but such replacements may still have a potential effect on the experimental results. In terms of E1, we conducted a linear regression and trend analysis on the mean of biases of all participants in each scale ratio because no significant differences existed among the biases of participants. Rather than investigating the biases of all individuals, we focused on the mean performance of the population, which was similar to the study of Harrison et al.~\cite{harrison2014ranking}. For E3, we had not yet obtained an accurate model to guide the bias correction. The main reason may be that the tested levels of point radii and scale ratios remained insufficient. Moreover, error bars in \autoref{fig6} are large because we scaled the range of $y$-axis to make the trends easy to read.

We recorded the time spent on each trial, but we did not perform any analysis on the time because the participants were allowed to repeatedly modify their answers until they were satisfied. No time limit was set on a trial. Moreover, we found in the significance analysis that most experimental results did not conform to a normal distribution. This may be caused by the small number of participants. In addition, \autoref{fig6} shows that the effect sizes of biases for numerosity seem substantial, and those for correlation and cluster separation seem small. However, effect sizes of biases for the three visual features could not be compared directly because of different perceptions and metrics.

\textbf{Potential future directions.} Further study on the breadth and depth can be conducted. 
For the former, we plan to investigate whether additional visual features can be affected by geometric scaling. We can also examine the influence of other distributions and visual channels on the bias. 
For the latter, we plan to explore the inner relations between the perceptions of high-level and low-level tasks. An example~\cite{ellis2018cognitive} shows why the cluster separation is perceptually biased from the visual density. We can also explore the relationships between features and biases. Experimental conditions, such as additional scale ratios and visual feature levels, should be enriched. An extensive range of participants and realistic multi-device environments should be involved. Our ultimate goal is to find a quantitative model to describe the bias, which can be used to formulate recommendations for automatic scaling.
\section{Conclusion}
In this study, we investigated the bias on the perceived visual features of scatterplots caused by geometric scaling. We proposed three hypotheses based on practical experience, and conducted a series of controlled experiments to test against the hypotheses. By analyzing the experimental results, we found that such a bias occurred and was linearly related to the scale ratio, thereby affecting the perception inconsistency of the scatterplots. We also found that no significant difference existed between the biases measured from normally and uniformly distributed scatterplots, and the bias could be partially corrected by changing the point radius. Our work is the first exploration of the bias caused by scatterplot scaling. We hope that this work will inspire other researchers to further study the perceptual process of scatterplot scaling, which should be a meaningful direction for enhancing the scatterplot design and promoting collaborative data exploration. We also expect that similar phenomena can be studied on other visualization technologies, such as bar and line charts, which should result in many interesting findings.

% \begin{equation}
% \sum_{j=1}^{z} j = \frac{z(z+1)}{2}
% \end{equation}

\acknowledgments{
We wish to thank all the anonymous reviewers for their valuable comments, and all the participants for their active participation. The work is supported by the National Key Research and Development Program of China (No. 2018YFB1700403) and the National Natural Science Foundation of China (No. 61772456, 61761136020, U1609217, 61672538 and 61872388).}

\bibliographystyle{abbrv-doi}

\bibliography{template}

\begin{thebibliography}{10}

\bibitem{alexander2017perceptual}
E.~Alexander, C.-C. Chang, M.~Shimabukuro, S.~Franconeri, C.~Collins, and
  M.~Gleicher.
\newblock Perceptual biases in font size as a data encoding.
\newblock {\em IEEE Transactions on Visualization and Computer Graphics},
  24(8):2397--2410, 2017.

\bibitem{allik1991occupancy}
J.~Allik and T.~Tuulmets.
\newblock {Occupancy model of perceived numerosity}.
\newblock {\em Perception \& Psychophysics}, 49(4):303--314, 1991.

\bibitem{andrews2010space}
C.~Andrews, A.~Endert, and C.~North.
\newblock {Space to Think: Large High-Resolution Displays for Sensemaking}.
\newblock In {\em Proceedings of the SIGCHI conference on human factors in
  computing systems}, pp. 55--64. ACM, 2010.

\bibitem{andrews2011information}
C.~Andrews, A.~Endert, B.~Yost, and C.~North.
\newblock {Information visualization on large, high-resolution displays:
  Issues, challenges, and opportunities}.
\newblock {\em Information Visualization}, 10(4):341--355, 2011.

\bibitem{anobile2014separate}
G.~Anobile, G.~M. Cicchini, and D.~C. Burr.
\newblock {Separate Mechanisms for Perception of Numerosity and Density}.
\newblock {\em Psychological Science}, 25(1):265--270, 2014.

\bibitem{anobile2015mechanisms}
G.~Anobile, M.~Turi, G.~M. Cicchini, and D.~C. Burr.
\newblock {Mechanisms for perception of numerosity or texture-density are
  governed by crowding-like effects}.
\newblock {\em Journal of Vision}, 15(5):4--4, 2015.

\bibitem{ardito2015interaction}
C.~Ardito, P.~Buono, M.~F. Costabile, and G.~Desolda.
\newblock {Interaction with Large Displays: A Survey}.
\newblock {\em ACM Computing Surveys (CSUR)}, 47(3):46, 2015.

\bibitem{behrisch2018quality}
M.~Behrisch, M.~Blumenschein, N.~W. Kim, L.~Shao, M.~El-Assady, J.~Fuchs,
  D.~Seebacher, A.~Diehl, U.~Brandes, H.~Pfister, et~al.
\newblock {Quality Metrics for Information Visualization}.
\newblock {\em Computer Graphics Forum}, 37(3):625--662, 2018.

\bibitem{bertini2011quality}
E.~Bertini, A.~Tatu, and D.~Keim.
\newblock {Quality Metrics in High-Dimensional Data Visualization: An Overview
  and Systematization}.
\newblock {\em IEEE Transactions on Visualization and Computer Graphics},
  17(12):2203--2212, 2011.

\bibitem{blascheck2019glanceable}
T.~Blascheck, L.~Besan{\c{c}}on, A.~Bezerianos, B.~Lee, and P.~Isenberg.
\newblock {Glanceable Visualization: Studies of Data Comparison Performance on
  Smartwatches}.
\newblock {\em IEEE Transactions on Visualization and Computer Graphics},
  25(1):630--640, 2019.

\bibitem{borkin2016beyond}
M.~A. Borkin, Z.~Bylinskii, N.~W. Kim, C.~M. Bainbridge, C.~S. Yeh, D.~Borkin,
  H.~Pfister, and A.~Oliva.
\newblock {Beyond Memorability: Visualization Recognition and Recall}.
\newblock {\em IEEE Transactions on Visualization and Computer Graphics},
  22(1):519--528, 2016.

\bibitem{brehmer2013multi}
M.~Brehmer and T.~Munzner.
\newblock {A Multi-Level Typology of Abstract Visualization Tasks}.
\newblock {\em IEEE Transactions on Visualization and Computer Graphics},
  19(12):2376--2385, 2013.

\bibitem{butscher2018clusters}
S.~Butscher, S.~Hubenschmid, J.~M{\"u}ller, J.~Fuchs, and H.~Reiterer.
\newblock {Clusters, Trends, and Outliers: How Immersive Technologies Can
  Facilitate the Collaborative Analysis of Multidimensional Data}.
\newblock In {\em Proceedings of the 2018 CHI Conference on Human Factors in
  Computing Systems}, pp. 90--100. ACM, 2018.

\bibitem{chan2010flow}
Y.-H. Chan, C.~D. Correa, and K.-L. Ma.
\newblock {Flow-based Scatterplots for Sensitivity Analysis}.
\newblock In {\em Proceedings of the 2010 IEEE Symposium on Visual analytics
  science and technology (VAST)}, pp. 43--50. IEEE, 2010.

\bibitem{chen2014visual}
H.~Chen, W.~Chen, H.~Mei, Z.~Liu, K.~Zhou, W.~Chen, W.~Gu, and K.-L. Ma.
\newblock Visual abstraction and exploration of multi-class scatterplots.
\newblock {\em IEEE Transactions on Visualization and Computer Graphics},
  20(12):1683--1692, 2014.

\bibitem{chittaro2006visualizing}
L.~Chittaro.
\newblock {Visualizing Information on Mobile Devices}.
\newblock {\em Computer}, 39(3):40--45, 2006.

\bibitem{chung2013comparison}
H.~Chung, S.~L. Chu, and C.~North.
\newblock {A comparison of Two Display Models for Collaborative Sensemaking}.
\newblock In {\em Proceedings of the 2nd ACM International Symposium on
  Pervasive Displays}, pp. 37--42. ACM, 2013.

\bibitem{cleveland1985elements}
W.~S. Cleveland and W.~S. Cleveland.
\newblock {\em {The Elements of Graphing Data}}, vol.~2.
\newblock Wadsworth Advanced Books and Software Monterey, CA, 1985.

\bibitem{cleveland1982variables}
W.~S. Cleveland, P.~Diaconis, and R.~McGill.
\newblock {Variables on Scatterplots Look More Highly Correlated When the
  Scales are Increased}.
\newblock {\em Science}, 216(4550):1138--1141, 1982.

\bibitem{cleveland1988shape}
W.~S. Cleveland, M.~E. McGill, and R.~McGill.
\newblock {The Shape Parameter of a Two-variable Graph}.
\newblock {\em Journal of the American Statistical Association},
  83(402):289--300, 1988.

\bibitem{conati2008exploring}
C.~Conati and H.~Maclaren.
\newblock {Exploring the Role of Individual Differences in Information
  Visualization}.
\newblock In {\em Proceedings of the working conference on Advanced visual
  interfaces}, pp. 199--206. ACM, 2008.

\bibitem{dakin2011common}
S.~C. Dakin, M.~S. Tibber, J.~A. Greenwood, M.~J. Morgan, et~al.
\newblock {A common visual metric for approximate number and density}.
\newblock {\em Proceedings of the National Academy of Sciences},
  108(49):19552--19557, 2011.

\bibitem{drucker2013touchviz}
S.~M. Drucker, D.~Fisher, R.~Sadana, J.~Herron, et~al.
\newblock {TouchViz: A Case Study Comparing Two Interfaces for Data Analytics
  on Tablets}.
\newblock In {\em Proceedings of the SIGCHI Conference on Human Factors in
  Computing Systems}, pp. 2301--2310. ACM, 2013.

\bibitem{ellis2018cognitive}
G.~Ellis.
\newblock {\em Cognitive Biases in Visualizations}.
\newblock Springer, 2018.

\bibitem{friendly2005early}
M.~Friendly and D.~Denis.
\newblock {The Early Origins and Development of The Scatterplot}.
\newblock {\em Journal of the History of the Behavioral Sciences},
  41(2):103--130, 2005.

\bibitem{garcia1998forced}
M.~A. Garc{\i}a-P{\'e}rez.
\newblock {Forced-choice staircases with fixed step sizes: asymptotic and
  small-sample properties}.
\newblock {\em Vision Research}, 38(12):1861--1881, 1998.

\bibitem{gleicher2013perception}
M.~Gleicher, M.~Correll, C.~Nothelfer, and S.~Franconeri.
\newblock {Perception of Average Value in Multiclass Scatterplots}.
\newblock {\em IEEE Transactions on Visualization and Computer Graphics},
  19(12):2316--2325, 2013.

\bibitem{gutwin2017peripheral}
C.~Gutwin, A.~Cockburn, and A.~Coveney.
\newblock {Peripheral Popout: The Influence of Visual Angle and Stimulus
  Intensity on Popout Effects}.
\newblock In {\em Proceedings of the 2017 CHI Conference on Human Factors in
  Computing Systems}, pp. 208--219. ACM, 2017.

\bibitem{gutwin2004interacting}
C.~Gutwin and C.~Fedak.
\newblock {Interacting with Big Interfaces on Small Screens: a Comparison of
  Fisheye, Zoom, and Panning Techniques}.
\newblock In {\em Proceedings of Graphics Interface 2004}, pp. 145--152.
  Canadian Human-Computer Communications Society, 2004.

\bibitem{harrison2015infographic}
L.~Harrison, K.~Reinecke, and R.~Chang.
\newblock {Infographic aesthetics: Designing for the first impression}.
\newblock In {\em Proceedings of the 33rd Annual ACM Conference on Human
  Factors in Computing Systems}, pp. 1187--1190. ACM, 2015.

\bibitem{harrison2014ranking}
L.~Harrison, F.~Yang, S.~Franconeri, and R.~Chang.
\newblock {Ranking Visualizations of Correlation Using Weber's Law}.
\newblock {\em IEEE Transactions on Visualization and Computer Graphics},
  20(12):1943--1952, 2014.

\bibitem{healey2012attention}
C.~Healey and J.~Enns.
\newblock {Attention and Visual Memory in Visualization and Computer Graphics}.
\newblock {\em IEEE Transactions on Visualization and Computer Graphics},
  18(7):1170--1188, 2012.

\bibitem{heer2006multi}
J.~Heer and M.~Agrawala.
\newblock {Multi-Scale Banking to 45 Degrees}.
\newblock {\em IEEE Transactions on Visualization and Computer Graphics},
  12(5):701--708, 2006.

\bibitem{jerit2012partisan}
J.~Jerit and J.~Barabas.
\newblock {Partisan Perceptual Bias and the Information Environment}.
\newblock {\em The Journal of Politics}, 74(3):672--684, 2012.

\bibitem{kandogan2012just}
E.~Kandogan.
\newblock {Just-in-time Annotation of Clusters, Outliers, and Trends in
  Point-based Data Visualizations}.
\newblock In {\em Proceedings of the 2012 IEEE Symposium on Visual Analytics
  Science and Technology (VAST)}, pp. 73--82. IEEE, 2012.

\bibitem{kay2016beyond}
M.~Kay and J.~Heer.
\newblock {Beyond Weber's Law: A Second Look at Ranking Visualizations of
  Correlation}.
\newblock {\em IEEE Transactions on Visualization and Computer Graphics},
  22(1):469--478, 2016.

\bibitem{kim2014colour}
S.~Kim, M.~Al-Haj, S.~Chen, S.~Fuller, U.~Jain, M.~Carrasco, and R.~Tannock.
\newblock {Colour vision in ADHD: Part 1-Testing the retinal dopaminergic
  hypothesis}.
\newblock {\em Behavioral and Brain Functions}, 10(1):38, 2014.

\bibitem{kwon2016study}
O.-H. Kwon, C.~Muelder, K.~Lee, and K.-L. Ma.
\newblock {A Study of Layout, Rendering, and Interaction Methods for Immersive
  Graph Visualization}.
\newblock {\em IEEE Transactions on Visualization and Computer Graphics},
  22(7):1802--1815, 2016.

\bibitem{li2010judging}
J.~Li, J.-B. Martens, and J.~J. Van~Wijk.
\newblock {Judging correlation from scatterplots and parallel coordinate
  plots}.
\newblock {\em Information Visualization}, 9(1):13--30, 2010.

\bibitem{li2010model}
J.~Li, J.~J. van Wijk, and J.-B. Martens.
\newblock {A Model of Symbol Lightness Discrimination in Sparse Scatterplots}.
\newblock In {\em Proceedings of the 2010 IEEE Symposium on Pacific
  Visualization (PacificVis)}, pp. 105--112. IEEE, 2010.

\bibitem{liao2017cluster}
H.~Liao, Y.~Wu, L.~Chen, and W.~Chen.
\newblock Cluster-based visual abstraction for multivariate scatterplots.
\newblock {\em IEEE Transactions on Visualization and Computer Graphics},
  24(9):2531--2545, 2017.

\bibitem{lv2018crowd}
P.~Lv, Z.~Zhang, C.~Li, Y.~Guo, B.~Zhou, and M.~Xu.
\newblock Crowd behavior evolution with emotional contagion in political
  rallies.
\newblock {\em IEEE Transactions on Computational Social Systems},
  6(2):377--386, 2018.

\bibitem{ma2018scatternet}
Y.~Ma, A.~K. Tung, W.~Wang, X.~Gao, Z.~Pan, and W.~Chen.
\newblock Scatternet: A deep subjective similarity model for visual analysis of
  scatterplots.
\newblock {\em IEEE Transactions on Visualization and Computer Graphics}, pp.
  1--1, 2018.

\bibitem{matejka2015dynamic}
J.~Matejka, F.~Anderson, and G.~Fitzmaurice.
\newblock {Dynamic Opacity Optimization for Scatter Plots}.
\newblock In {\em Proceedings of the 33rd Annual ACM Conference on Human
  Factors in Computing Systems}, pp. 2707--2710. ACM, 2015.

\bibitem{mei2018design}
H.~Mei, Y.~Ma, Y.~Wei, and W.~Chen.
\newblock The design space of construction tools for information visualization:
  A survey.
\newblock {\em Journal of Visual Languages \& Computing}, 44:120--132, 2018.

\bibitem{ni2006survey}
T.~Ni, G.~S. Schmidt, O.~G. Staadt, M.~A. Livingston, R.~Ball, and R.~May.
\newblock {A Survey of Large High-Resolution Display Technologies, Techniques,
  and Applications}.
\newblock In {\em Proceedings of the 2006 IEEE Conference on Virtual Reality},
  pp. 223--236. IEEE, 2006.

\bibitem{ondov2019face}
B.~Ondov, N.~Jardine, N.~Elmqvist, and S.~Franconeri.
\newblock {Face to Face: Evaluating Visual Comparison}.
\newblock {\em IEEE Transactions on Visualization and Computer Graphics},
  25(1):861--871, 2019.

\bibitem{peng2004clutter}
W.~Peng, M.~O. Ward, and E.~A. Rundensteiner.
\newblock {Clutter Reduction in Multi-Dimensional Data Visualization Using
  Dimension Reordering}.
\newblock In {\em Proceedings of the 2004 IEEE Symposium on Information
  Visualization}, pp. 89--96. IEEE, 2004.

\bibitem{reach2019smooth}
A.~M. Reach and C.~North.
\newblock {Smooth, Efficient, and Interruptible Zooming and Panning}.
\newblock {\em IEEE Transactions on Visualization and Computer Graphics},
  25(2):1421--1434, 2019.

\bibitem{reda2015effects}
K.~Reda, A.~E. Johnson, M.~E. Papka, and J.~Leigh.
\newblock {Effects of Display Size and Resolution on User Behavior and Insight
  Acquisition in Visual Exploration}.
\newblock In {\em Proceedings of the 33rd Annual ACM Conference on Human
  Factors in Computing Systems}, pp. 2759--2768. ACM, 2015.

\bibitem{rensink2017nature}
R.~A. Rensink.
\newblock {The nature of correlation perception in scatterplots}.
\newblock {\em Psychonomic Bulletin \& Review}, 24(3):776--797, 2017.

\bibitem{rensink2010perception}
R.~A. Rensink and G.~Baldridge.
\newblock {The Perception of Correlation in Scatterplots}.
\newblock {\em Computer Graphics Forum}, 29(3):1203--1210, 2010.

\bibitem{ross2010vision}
J.~Ross and D.~C. Burr.
\newblock {Vision senses number directly}.
\newblock {\em Journal of vision}, 10(2):10--10, 2010.

\bibitem{saket2016beyond}
B.~Saket, A.~Endert, and J.~Stasko.
\newblock {Beyond Usability and Performance: A Review of User
  Experience-focused Evaluations in Visualization}.
\newblock In {\em Proceedings of the Sixth Workshop on Beyond Time and Errors
  on Novel Evaluation Methods for Visualization}, pp. 133--142. ACM, 2016.

\bibitem{saket2018evaluating}
B.~Saket, A.~Srinivasan, E.~D. Ragan, and A.~Endert.
\newblock {Evaluating Interactive Graphical Encodings for Data Visualization}.
\newblock {\em IEEE Transactions on Visualization and Computer Graphics},
  24(3):1316--1330, 2018.

\bibitem{sarikaya2018scatterplots}
A.~Sarikaya and M.~Gleicher.
\newblock {Scatterplots: Tasks, Data, and Designs}.
\newblock {\em IEEE Transactions on Visualization and Computer Graphics},
  24(1):402--412, 2018.

\bibitem{schulz2013design}
H.-J. Schulz, T.~Nocke, M.~Heitzler, and H.~Schumann.
\newblock {A Design Space of Visualization Tasks}.
\newblock {\em IEEE Transactions on Visualization and Computer Graphics},
  19(12):2366--2375, 2013.

\bibitem{sedlmair2015data}
M.~Sedlmair and M.~Aupetit.
\newblock {Data-driven Evaluation of Visual Quality Measures}.
\newblock {\em Computer Graphics Forum}, 34(3):201--210, 2015.

\bibitem{sedlmair2013empirical}
M.~Sedlmair, T.~Munzner, and M.~Tory.
\newblock {Empirical Guidance on Scatterplot and Dimension Reduction Technique
  Choices}.
\newblock {\em IEEE Transactions on Visualization and Computer Graphics},
  19(12):2634--2643, 2013.

\bibitem{sedlmair2012taxonomy}
M.~Sedlmair, A.~Tatu, T.~Munzner, and M.~Tory.
\newblock {A Taxonomy of Visual Cluster Separation Factors}.
\newblock {\em Computer Graphics Forum}, 31(3):1335--1344, 2012.

\bibitem{shi2018novel}
Y.~Shi, Y.~Zhao, F.~Zhou, R.~Shi, and Y.~Zhang.
\newblock A novel radial visualization of intrusion detection alerts.
\newblock {\em IEEE Computer Graphics and Applications}, 38(6):83--95, 2018.

\bibitem{sips2009selecting}
M.~Sips, B.~Neubert, J.~P. Lewis, and P.~Hanrahan.
\newblock {Selecting good views of high-dimensional data using class
  consistency}.
\newblock {\em Computer Graphics Forum}, 28(3):831--838, 2009.

\bibitem{sophian2008people}
C.~Sophian and Y.~Chu.
\newblock {How do people apprehend large numerosities?}
\newblock {\em Cognition}, 107(2):460--478, 2008.

\bibitem{steichen2013user}
B.~Steichen, G.~Carenini, and C.~Conati.
\newblock {User-adaptive information visualization: using eye gaze data to
  infer visualization tasks and user cognitive abilities}.
\newblock In {\em Proceedings of the 2013 international conference on
  Intelligent user interfaces}, pp. 317--328. ACM, 2013.

\bibitem{stevens1986psychophysics}
S.~S. Stevens and G.~Stevens.
\newblock {Psychophysics: Introduction to its perceptual, neural, and social
  prospects. New Brunswick}, 1986.

\bibitem{szafir2018modeling}
D.~A. Szafir.
\newblock {Modeling Color Difference for Visualization Design}.
\newblock {\em IEEE Transactions on Visualization and Computer Graphics},
  24(1):392--401, 2018.

\bibitem{tatu2009combining}
A.~Tatu, G.~Albuquerque, M.~Eisemann, J.~Schneidewind, H.~Theisel, M.~Magnork,
  and D.~Keim.
\newblock {Combining automated analysis and visualization techniques for
  effective exploration of high-dimensional data}.
\newblock In {\em Proceedings of the 2009 IEEE Symposium on Visual Analytics
  Science and Technology (VAST)}, pp. 59--66. IEEE, 2009.

\bibitem{tatu2010visual}
A.~Tatu, P.~Bak, E.~Bertini, D.~Keim, and J.~Schneidewind.
\newblock {Visual quality metrics and human perception: an initial study on 2D
  projections of large multidimensional data}.
\newblock In {\em Proceedings of the International Conference on Advanced
  Visual Interfaces}, pp. 49--56. ACM, 2010.

\bibitem{tibber2012number}
M.~S. Tibber, J.~A. Greenwood, and S.~C. Dakin.
\newblock {Number and density discrimination rely on a common metric: Similar
  psychophysical effects of size, contrast, and divided attention}.
\newblock {\em Journal of Vision}, 12(6):1--19, 2012.

\bibitem{tibber2013sensitivity}
M.~S. Tibber, G.~S. Manasseh, R.~C. Clarke, G.~Gagin, S.~N. Swanbeck,
  B.~Butterworth, R.~B. Lotto, and S.~C. Dakin.
\newblock {Sensitivity to numerosity is not a unique visuospatial
  psychophysical predictor of mathematical ability}.
\newblock {\em Vision Research}, 89:1--9, 2013.

\bibitem{tokita2010might}
M.~Tokita and A.~Ishiguchi.
\newblock {How might the discrepancy in the effects of perceptual variables on
  numerosity judgment be reconciled?}
\newblock {\em Attention, Perception, \& Psychophysics}, 72(7):1839--1853,
  2010.

\bibitem{tory2007spatialization}
M.~Tory, D.~Sprague, F.~Wu, W.~Y. So, and T.~Munzner.
\newblock {Spatialization Design: Comparing Points and Landscapes}.
\newblock {\em IEEE Transactions on Visualization and Computer Graphics},
  13(6):1262--1269, 2007.

\bibitem{valdez2018priming}
A.~C. Valdez, M.~Ziefle, and M.~Sedlmair.
\newblock {Priming and Anchoring Effects in Visualization}.
\newblock {\em IEEE Transactions on Visualization and Computer Graphics},
  24(1):584--594, 2018.

\bibitem{wang2018shadow}
H.~Wang, M.~Xu, F.~Zhu, Z.~Deng, Y.~Li, and B.~Zhou.
\newblock Shadow traffic: A unified model for abnormal traffic behavior
  simulation.
\newblock {\em Computers \& Graphics}, 70:235--241, 2018.

\bibitem{wang2019optimizing}
Y.~Wang, X.~Chen, T.~Ge, C.~Bao, M.~Sedlmair, C.-W. Fu, O.~Deussen, and
  B.~Chen.
\newblock Optimizing color assignment for perception of class separability in
  multiclass scatterplots.
\newblock {\em IEEE Transactions on Visualization and Computer Graphics},
  25(1):820--829, 2019.

\bibitem{Wang2019Image}
Y.~Wang, Z.~Wang, C.~W. Fu, H.~Schmauder, O.~Deussen, and D.~Weiskopf.
\newblock Image-based aspect ratio selection.
\newblock {\em IEEE Transactions on Visualization and Computer Graphics},
  25(1):840--849, 2019.

\bibitem{ware2012information}
C.~Ware.
\newblock {\em {Information Visualization: Perception for Design}}.
\newblock Elsevier, 2012.

\bibitem{xia2018exploring}
J.~Xia, L.~Gao, K.~Kong, Y.~Zhao, Y.~Chen, X.~Kui, and Y.~Liang.
\newblock Exploring linear projections for revealing clusters, outliers, and
  trends in subsets of multi-dimensional datasets.
\newblock {\em Journal of Visual Languages \& Computing}, 48:52--60, 2018.

\bibitem{xia2017ldsscanner}
J.~Xia, F.~Ye, W.~Chen, Y.~Wang, W.~Chen, Y.~Ma, and A.~K. Tung.
\newblock Ldsscanner: exploratory analysis of low-dimensional structures in
  high-dimensional datasets.
\newblock {\em IEEE Transactions on Visualization and Computer Graphics},
  24(1):236--245, 2017.

\bibitem{xu2017efficient}
M.~Xu, C.~Li, P.~Lv, N.~Lin, R.~Hou, and B.~Zhou.
\newblock An efficient method of crowd aggregation computation in public areas.
\newblock {\em IEEE Transactions on Circuits and Systems for Video Technology},
  28(10):2814--2825, 2017.

\bibitem{yang2019correlation}
F.~Yang, L.~T. Harrison, R.~A. Rensink, S.~L. Franconeri, and R.~Chang.
\newblock {Correlation Judgment and Visualization Features: A Comparative
  Study}.
\newblock {\em IEEE Transactions on Visualization and Computer Graphics},
  25(3):1474--1488, 2019.

\bibitem{yost2006perceptual}
B.~Yost and C.~North.
\newblock {The Perceptual Scalability of Visualization}.
\newblock {\em IEEE Transactions on Visualization and Computer Graphics},
  12(5):837--844, 2006.

\bibitem{zhao2019evaluating}
Y.~Zhao, F.~Luo, M.~Chen, Y.~Wang, J.~Xia, F.~Zhou, Y.~Wang, Y.~Chen, and
  W.~Chen.
\newblock {Evaluating Multi-Dimensional Visualizations for Understanding Fuzzy
  Clusters}.
\newblock {\em IEEE Transactions on Visualization and Computer Graphics},
  25(1):12--21, 2019.

\bibitem{zhou2019survey}
F.~Zhou, X.~Lin, C.~Liu, Y.~Zhao, P.~Xu, L.~Ren, T.~Xue, and L.~Ren.
\newblock A survey of visualization for smart manufacturing.
\newblock {\em Journal of Visualization}, 22(2):419--435, 2019.

\bibitem{zhou2018visual}
Z.~Zhou, L.~Meng, C.~Tang, Y.~Zhao, Z.~Guo, M.~Hu, and W.~Chen.
\newblock Visual abstraction of large scale geospatial origin-destination
  movement data.
\newblock {\em IEEE Transactions on Visualization and Computer Graphics},
  25(1):43--53, 2018.

\bibitem{zimmermann2016numerosity}
E.~Zimmermann and G.~R. Fink.
\newblock {Numerosity perception after size adaptation}.
\newblock {\em Scientific Reports}, 6(1):32810--32817, 2016.

\end{thebibliography}
\end{document}